\documentclass[12pt]{article}

\usepackage[english]{babel}
\usepackage{amssymb,slashed,latexsym,amsmath,multirow}
\usepackage[font=small,format=plain,labelfont=bf,width=15cm]{caption}

\newcommand{\Symp}{\mathop{\rm {}Sp}}
\newcommand{\ft}[2]{{\textstyle\frac{#1}{#2}}}
\newcommand{\lc}{{|\!\!|\!\!\lceil}}
\newcommand{\rf}{{\rfloor\!\!|\!\!|}}
\newcommand{\VSone}{\widetilde{\boldsymbol{V}}_{\!\!\!\!\boldsymbol{(1)}}}
\newcommand{\ZSone}{\widetilde{\boldsymbol{Z}}_{\!\!\!\boldsymbol{(1)}}}
\newcommand{\VStwo}{\widetilde{\boldsymbol{V}}_{\!\!\!\!\boldsymbol{(2)}}}
\newcommand{\ZStwo}{\widetilde{\boldsymbol{Z}}_{\!\!\!\boldsymbol{(2)}}}

\newcommand{\RS}{\widetilde{\boldsymbol{R}}}
\newcommand{\R}{\boldsymbol{R}}
\newcommand{\Zone}{{\boldsymbol{Z}}_{\!\!\boldsymbol{(1)}}}
\newcommand{\Ztwo}{{\boldsymbol{Z}}_{\!\!\boldsymbol{(2)}}}

\newcommand{\TS}{\widetilde{\boldsymbol{T}}}
\newcommand{\T}{\boldsymbol{T}}
\newcommand{\ES}{\widetilde{\boldsymbol{E}}}

\newsavebox{\uuunit}
\sbox{\uuunit}
    {\setlength{\unitlength}{0.825em}
     \begin{picture}(0.6,0.7)
        \thinlines
        \put(0,0){\line(1,0){0.5}}
        \put(0.15,0){\line(0,1){0.7}}
        \put(0.35,0){\line(0,1){0.8}}
       \multiput(0.3,0.8)(-0.04,-0.02){12}{\rule{0.5pt}{0.5pt}}
     \end {picture}}

\csname @addtoreset\endcsname{equation}{section}

\begin{document}
\begin{titlepage}
\begin{flushright}
KUL-TF-10/01\\
arXiv:1001.2560
\end{flushright}
\vspace{.5cm}
\begin{center}
\baselineskip=16pt {\LARGE    Generalized gaugings  \\
\vskip 0.2cm and the field-antifield formalism
}\\
\vfill
{\large Frederik Coomans$^{\dag}$, Jan De Rydt$^{\dag}$, Antoine Van Proeyen$^{\dag}$
  } \\
\vfill
{\small Instituut voor Theoretische Fysica, Katholieke Universiteit Leuven,\\
       Celestijnenlaan 200D B-3001 Leuven, Belgium.
\\ \vspace{6pt}
 }
\end{center}
\vfill
\begin{center}
{\bf Abstract}
\end{center}
{\small We discuss the algebra of general gauge theories that are described by the embedding tensor
 formalism. We compare the gauge transformations dependent and
 independent of an invariant action, and argue that the generic transformations lead to an infinitely reducible algebra. We connect the embedding tensor
 formalism to the field-antifield (or Batalin-Vilkovisky) formalism,
 which is the most general formulation known for general gauge theories
 and their quantization. The structure
 equations of the embedding tensor formalism are included in the master equation of the field-antifield formalism. }

 \vfill

\hrule width 3.cm \vspace{2mm}{\footnotesize \noindent $^{\dag}$e-mails:
\{Frederik.Coomans, Jan.DeRydt, Antoine.VanProeyen\}@fys.kuleuven.be}
\end{titlepage}
\addtocounter{page}{1}
\tableofcontents{}
\newpage

\section{Introduction}
Supergravity theories exist in many varieties. Their properties depend primarily on the choice of spacetime dimension
$D$ and the number of supercharges ${\cal N}$.
For some of the values of $D$ and ${\cal N}$ one still has the freedom to decide which type of matter multiplets can
be added. Over the years, the various possibilities --called `basic supergravities'-- have been classified and they are well understood.
However, when the field content is fixed after the above steps, there are
still different theories possible. The simplest example being the choice
of the K{\"a}hler and superpotential in $D=4$, ${\cal N}=1$ supergravity.
Other examples of
`deformations' consist in the coupling to Yang-Mills type gauge groups
(see for example \cite{de_Wit:1982ig}) or the introduction of
mass-parameters \cite{Romans:1985tz}. Recently, a powerful technique has
been developed to systematically study and classify all these
deformations. This is the so-called embedding tensor formalism which was
first introduced in $D=3$ in \cite{Nicolai:2000sc,Nicolai:2001sv} and later
extended to higher dimensions in
\cite{deWit:2002vt,deWit:2004nw,deWit:2005hv,deWit:2005ub}. This
formalism is not restricted to supergravity theories, despite the fact
that its main application is in this context. The progress is rather
related to a better understanding of all the possibilities of gauge
groups and their coupling to all the fields in the theories.

The construction of the embedding tensor formalism starts from an undeformed theory --such as the basic supergravities--
with a rigid symmetry group $G_{\rm rigid}$ and generators $\delta_\alpha$, $\alpha=1,\ldots,{\rm dim}(\mathfrak{g}_{\rm rigid})$. The precise matter content of the undeformed
theory is not relevant for our purposes, although it is important to highlight the vector fields. These will be denoted by $A_\mu{}^M$ and transform under the rigid symmetry group with certain matrix multiplications. Once these ingredients are known, one introduces an embedding tensor, denoted by $\Theta_M{}^\alpha$, which selects a linear
combination of the rigid symmetry generators, and promotes them to local transformations, $\delta_M$:
\begin{equation}\label{embedding}
 \delta_M \equiv \sum_\alpha \Theta_M{}^\alpha \delta_\alpha\,.
\end{equation}
The number of independent generators $\delta_M$ is less than or equal to the number of vectors $A_\mu{}^M$. But since there are always Abelian transformations that work on the vectors with a $\partial_{\mu}\Lambda^M$ term, the total number of local generators is equal to the number of vectors. The complete set of generators forms a basis for the gauge algebra, denoted by ${\mathfrak g}_{\rm gauge}$, and $M=1,\,\hdots,\, {\rm dim}(\mathfrak{g}_{\rm gauge})$.
Closure of this gauge algebra requires that the commutator of two transformations is again a linear combination of transformations:
\begin{equation}\label{gaugealgebra}
 [\delta_M,\delta_N]=f_{MN}{}^P \delta_P\,,
\end{equation}
with $f_{MN}{}^P$ the structure constants.
In particular, equation (\ref{gaugealgebra}) should hold for the gauge generators in the vector representation. Clearly, this can only be true if we impose a constraint on the embedding tensor, which is called the closure constraint.
However, this constraint does not guarantee the validity of the Jacobi identities and as such, extra ingredients are necessary in order to restore consistency of the
construction. First, one has to extend the original gauge algebra $\mathfrak{g}_{\rm gauge}$ with new local transformations.\footnote{In section \ref{s:closure} we will see that we need three types of local transformations in total. They correspond to electromagnetic gauge parameters for the 1-forms, and gauge and shift
parameters for the antisymmetric 2-forms.} Part of these new
transformations can be used to gauge away the directions in the original algebra that violate the Jacobi identities. Second, in order to restore the right number of
degrees of freedom, one has to introduce new $2$-form tensor fields, denoted by $B_{\mu\nu}{}^{MN}$. In the end all these elements combine to form a consistent gauge structure, whose properties we will attempt to clarify in this paper.

Before we go deeper into the details of the gauge structure, let us remark that we have not introduced any dynamics for the fields so far. Nevertheless, the formalism that we discussed can be implemented into a framework with an action. For example, this has been done in \cite{deWit:2005ub} for $D=4$. However, this implementation requires a modification of the gauge structure that we have outlined above. First, one has to impose an extra constraint --called linear constraint-- on the embedding tensor, which guarantees the gauge invariance of the action. Once this constraint and the closure constraint are satisfied, the entries of the embedding tensor are nothing else but the deformation parameters
in the theory. Together with the rigid symmetry group $G_{\rm rigid}$, they determine all the gauge couplings in the action. The second modification is a change in the original gauge transformations of the $2$-forms. They need to be supplemented by extra terms, which will be denoted by $\Delta B$ and are characterized by a dependence on the matter fields (such as scalars) in the action.

To summarize, Table \ref{table1} emphasizes the ingredients that we have introduced so far.
\begin{table}[t]\begin{center}
\renewcommand{\arraystretch}{1.5}
\begin{tabular}{|l|l|l|}\hline
 & \multicolumn{1}{c|}{embedding tensor formalism}  & \multicolumn{1}{c|}{embedding tensor formalism} \vspace{-3mm}\\
 &\multicolumn{1}{c|}{without an action} &\multicolumn{1}{c|}{with an action}\\
 \hline constraints on $\Theta_M{}^\alpha$: & closure constraint & closure and linear constraint\\
 field content: & $A_{\mu}{}^M,B_{\mu\nu}{}^{MN}$ & $A_{\mu}{}^M,B_{\mu\nu}{}^{MN}$\\
 gauge transformations: & 3 types, $\Delta B=0$ & 3 types, $\Delta B\neq0$\\\hline
\end{tabular}\caption{\label{table1} Schematic overview and ingredients of the embedding tensor formalism.}\end{center}\end{table}
We note that the case without an action can be extended beyond the $2$-tensors, by adding extra $p$-forms ($p > 2$) and extra gauge transformations. This construction leads to the so-called `tensor hierarchy' \cite{deWit:2005hv,deWit:2008ta,deWit:2008gc,deWit:2009zv,Bergshoeff:2009ph}. The hierarchy can also be embedded into the
framework of an action, leading to an extension of the last column in Table \ref{table1}. More precisely, the de-forms ($p=D-1$ in $D$ dimensions) and top-forms ($p=D$) appear in the action as Lagrange multipliers for the constraints on the embedding tensor.

Although the full hierarchy has an intricate gauge structure with interesting properties, in this text we will only consider its truncation to the $p \leq 2$-forms. More precisely, we study the gauge algebra structure of the $D=4$ embedding tensor formalism up to $1$- and $2$-forms, both with and without a Lagrangian description. We obtain the following results:
\begin{enumerate}
  \item In both cases the algebra is reducible, which means that the 3 types of gauge transformations are not independent.
  This dependence is characterized by the so-called zero modes of the
  theory. In general, the system even turns out to be higher stage
  reducible; there exist zero modes for the zero modes etc. It is not
  clear if these higher stage zero modes break down after a finite number of steps, i.e. whether the
  algebra is finitely reducible.
  \item The algebra has a closed form if one starts from the symmetry transformations without specifying an action. On the other hand, once a particular form
for the action is introduced, the algebra of the modified gauge transformations (i.e. $\Delta B\neq 0$) is open. This means that the commutator of two modified gauge transformations only closes up to terms that are proportional to the equations of motion.
 \item In both cases the gauge algebra has a soft form, i.e. the `structure constants' are not really constant,
  but they depend on the fields in the theory.
\end{enumerate}
Hence, we find that the algebra of symmetries has a structure that is very involved. This brings us to the second purpose of this paper, which is to provide a more concise framework
for these complicated properties. This will be achieved through a new formulation that connects the embedding tensor formalism to the Batalin-Vilkovisky (or field-antifield) formalism.

The Batalin-Vilkovisky (BV) formalism \cite{Batalin:1984jr,Henneaux:1990jq,Gomis:1995he} was originally constructed as a method to quantize a broad class of gauged field theories. This quantization is not always straightforward, due to certain properties of the gauge algebra. A first complication arises when one considers non-Abelian gauge groups, for which it is not easy to find a gauge fixing procedure that preserves the symmetries of the system. A way to overcome this problem is provided by the Faddeev-Popov method, which introduces unphysical ghost fields to compensate for the gauge degrees of freedom. However, the Faddeev-Popov formalism does not directly apply to theories with a more complicated gauge structure, such
as systems with a reducible, soft or open gauge algebra. These are exactly the properties that we encountered in the algebra corresponding with the embedding tensor formalism. They also arise naturally in supergravity constructions, due to the presence of higher order
form fields, the use of non-supersymmetric gauges (such as the
Wess-Zumino gauge), the presence of off-shell multiplets, etc. In order
to deal with these complicated properties in a quantization procedure, it
is necessary to add extra ingredients to the Faddeev-Popov method, such
as ghosts for ghosts that compensate for the zero modes mentioned above. This extension of the Faddeev-Popov method is exactly given by the BV formalism;
it provides all necessary ingredients to quantize a class of generic gauge theories, having the properties that we outlined above.

Apart from being a mechanism for quantization, the BV formalism is also a powerful tool to describe \textit{classical} theories with complicated gauge structures. The reason is that all the properties (soft, open, reducible, $\ldots$) of the classical theory are captured by the so-called `extended action' of the BV formalism. This action is an extension of the usual Faddeev-Popov action and has to satisfy several
conditions, one of which is the `master equation'.\footnote{In the case of the Faddeev-Popov construction, the master equation reduces to the well-known BRST symmetry of the gauge fixed action.} In general, the master equation guarantees that the extended action is built up by tensors that determine the precise form of the algebra: gauge generators, structure functions, zero modes, etc. It is exactly this simple formulation that we will exploit in this text.

Since the complications that arose in the embedding tensor formalism (the fact that we are dealing with a soft, reducible and open algebra) are exactly those for which the BV formalism was designed, we can try to reformulate it in terms of the BV formalism. We recall here the basic ingredients that we need:
\begin{enumerate}
  \item For every gauge parameter in the algebra, we introduce a ghost field. For every zero mode, we introduce a ghost for ghost field, etc. \label{ghosts}
  \item Each of the fields in \ref{ghosts}. gets a corresponding antifield which should be regarded as a mathematical tool to set up the formalism.
  \item We construct the extended action as an expansion in the antifields and impose the master equation.
\end{enumerate}
We find that, at `zeroth order' in the antifields, the extended action is equal to the classical action. If, as mentioned above, we consider the gauge algebra in
the absence of a classical action then the extended action has no zeroth order part.
All higher order terms depend on tensors that reflect the gauge structure of the vector and tensor fields. For example, at first order we find the gauge generators,
second order contains the structure functions, etc. We will restrict our analysis to second order since this is enough to incorporate the most important properties of the algebra. We also expect that this investigation can be continued to arbitrary high orders in the antifields.

To summarize, we will motivate that a lot of the structure relations that appear in the different papers on the embedding formalism get unified in the master equation of the BV extended action. This result can be helpful in the future to gain more insight into the gauge structure at higher orders in the antifields. Moreover, due to our embedding in the BV formalism, we now have all the tools available to initiate the quantization of generic gauged field theories.

The outline of the article is as follows. In section \ref{s:embedding} we review the basic ingredients of the embedding tensor formalism. We will introduce the different tensor fields and their gauge transformations, together with a Lagrangian description. In section \ref{s:gaugealgebra}, the gauge structure of this formalism will be investigated and we conclude that its properties are exactly those for which the field-antifield formalism was constructed. This observation will be exploited in section \ref{s:BV}, where we reformulate the embedding tensor formalism in terms of the field-antifield formalism. Finally, in section \ref{s:conclusions} we conclude with a short summary and an outlook.

\section{The embedding tensor formalism \label{s:embedding}}
We recall from the introduction that possible gaugings of field theories with vectors can be classified using the embedding tensor formalism. This formalism has a universal structure that is independent of the number of spacetime dimensions or extra properties such as (local) supersymmetry. The only input that is needed is the rigid symmetry group of the undeformed theory, and its (vector) field content. In this section, we review the basic ingredients of the embedding tensor formalism, and pave the way for a more thorough discussion of the particular gauge structures that arise. To simplify matters, we will work in 4 dimensions from now on.

\subsection{General structure \label{s:structure}}
The basic supergravities in $D=4$ have a rigid symmetry group that is contained in the product of the symplectic duality group and the isometry group of the scalar manifold:
\begin{equation}
 G_{\rm rigid} \subseteq \Symp(2n_V+2,\mathbb{R})\times {\rm Iso}({\mathcal M}_{\rm scalar})\,.
\end{equation}
The symplectic transformations in this product work on the vector sector, with $n_V$ the number of vector multiplets. In general, the vector fields in the theory will be denoted by $A_\mu{}^M$ with $M$ a symplectic index that takes $2n_V+2$ values. They can be split into an electric and a magnetic part, usually denoted by an upper and lower index $\Lambda$ respectively: $A_\mu{}^M=(A_\mu{}^\Lambda, A_{\mu \Lambda})$. Under the rigid symmetry group, the vectors transform as follows:
\begin{equation}\label{globaltransfoA}
 \delta(\lambda) A_\mu{}^M = \lambda^\alpha\delta_\alpha A_\mu{}^M = - \lambda^\alpha (t_\alpha)_N{}^M A_\mu{}^N\,,
\end{equation}
where the $(t_\alpha)_N{}^M$ are the rigid symmetry generators in the vector representation.
The matrices $(t_\alpha)_N{}^M$ satisfy the symplectic condition $(t_\alpha)_{[N}{}^M \Omega_{P]M}=0$, with $\Omega_{PM}$ the symplectic metric.

Once these features are known, one can proceed with the construction of the embedding tensor formalism. Its main ideas are
\begin{enumerate}
 \item the embedding of the \textit{local} symmetries in (a subgroup of) the rigid invariance group,
\item the use of (some of) the vectors $A_\mu{}^M$ as corresponding gauge fields.
\end{enumerate}
This is achieved through the
introduction of a tensor $\Theta_M{}^\alpha$, which selects linear combinations of global symmetry generators to become local gauge transformations;
recall equation (\ref{embedding}).
The tensor $\Theta_M{}^\alpha$ has two indices; the upper index $\alpha$ labels the generators of the rigid symmetry group, and the lower index $M$ is the symplectic vector index that now also labels the local generators.

Once the gauge generators have been selected through the choice of an embedding tensor, one proceeds as usual by introducing a local parameter, $\Lambda^M(x)$, for every generator and by constructing
covariant derivatives. The gauge transformations of the vectors are
\begin{equation}\label{localtransfoA}
 \delta(\Lambda)A_\mu{}^M = \partial_\mu\Lambda^M + A_\mu{}^N X_{NP}{}^M \Lambda^P \equiv D_\mu \Lambda^M\,,
\end{equation}
with $X_{NP}{}^M \equiv \Theta_N{}^\alpha (t_\alpha)_P{}^M$. Covariant derivatives take the following form:
\begin{equation}
 D_\mu \equiv \partial_\mu - A_\mu{}^M \delta_M=\partial_\mu - A_\mu{}^M \Theta_M{}^\alpha \delta_\alpha\,.
\end{equation}
For the construction of gauge theories in terms of a regular Lie algebra, this would be the end. However, in the case of the embedding tensor formalism, the ingredients that we have introduced are not sufficient to construct a gauge covariant theory. There are certain complications that arise because of the special properties of the matrices $(X_M)_N{}^P=X_{MN}{}^P$. For example, the usual definition of the field strengths,
 \begin{equation}\label{ordinaryF}
   {\cal F}_{\mu\nu}{}^M \equiv 2 \partial_{[\mu}A_{\nu]}{}^M + X_{[NP]}{}^M A_\mu{}^N A_\nu{}^P\,,
 \end{equation}
is not covariant under the transformations (\ref{localtransfoA}). These problems and corresponding solutions will be discussed now.

Let us start by pointing out that a consistent gauging requires the imposition of the so-called `closure constraint' on the embedding tensor. It has the following form:
 \begin{equation}\label{closureconstr}
Q_{MN}{}^{\alpha} \equiv f_{\beta\gamma}{}^{\alpha}\Theta_M{}^{\beta}\Theta_N{}^{\gamma}+(t_{\beta})_N{}^Q\Theta_M{}^{\beta}\Theta_Q{}^{\alpha}=0\,,
 \end{equation}
where $f_{\alpha\beta}{}^{\gamma}$ are the structure constants of the rigid symmetry algebra, $[\delta_{\alpha},\delta_{\beta}]=f_{\alpha\beta}{}^{\gamma}\delta_{\gamma}$.
This constraint is equivalent to demanding gauge invariance of the embedding tensor, since $Q_{MN}{}^{\alpha}=\delta_M \Theta_N{}^\alpha$. It also implies that
\begin{equation}\label{closurealgebra}
 [X_M,X_N]=-X_{MN}{}^P X_P\,,
\end{equation}
and therefore guarantees the closure of the gauge algebra, with $X_{MN}{}^P$ as its generalized structure constants. From their definition below (\ref{localtransfoA}), it is clear that the $X_{MN}{}^P$ are in general not antisymmetric in $[MN]$. However, the left hand side of (\ref{closurealgebra}) is antisymmetric in $[MN]$, and therefore, so should be the right hand side. This means that we have to impose
\begin{equation}\label{ZTheta}
 Y^P{}_{MN} X_P = 0\,,\qquad \mbox{with }\;  Y^P{}_{MN}\equiv X_{(MN)}{}^P\,.
\end{equation}
 Thus the symmetric part of $X_{MN}{}^P$ only vanishes upon contraction with the embedding tensor, but is not zero in itself. This signals a difference with ordinary gauge groups, where the structure constants are antisymmetric and satisfy the Jacobi identity. In our case, the Jacobi identity is violated by terms that are proportional to $Y^P{}_{MN}$:
\begin{equation} \label{Jacobiidentity}
  X_{[MN]}{}^P X_{[QP]}{}^R + \mbox{cyclic}= -\frac{1}{3}\left(X_{[MN]}{}^P Y^R{}_{QP} + \mbox{cyclic}\right)\,.
\end{equation}
This in turn requires several modifications to the structure that we just outlined.

The first step towards a solution is the introduction of extra gauge transformations for the vector fields,
accompanied by new local parameters $\Xi_\mu{}^{NP}(x)$:
\begin{equation}\label{deltaA}
 \delta(\Lambda) A_\mu{}^M\;\rightarrow\;\delta(\Lambda,\Xi) A_\mu{}^M = \delta(\Lambda) A_\mu{}^M + \delta(\Xi) A_\mu{}^M,
\end{equation}
where
\begin{equation}
\delta(\Xi) A_\mu{}^M = - Y^M{}_{NP} \Xi_\mu{}^{NP}\,.
\end{equation}
The new transformations are proportional to the symmetric part of $X_{MN}{}^P$, and are introduced to gauge away the directions in the algebra that violate
the Jacobi identity.

As a side remark, we note that also the gauge algebra provides evidence for the necessity of extra transformations $\delta(\Xi)$. We will give a more detailed discussion of the algebra later, but for now, let us clarify our point and compute the commutator of two $\delta(\Lambda)$-transformations on the gauge fields:
\begin{equation}
 \label{algebraLLA} [\delta(\Lambda_1),\delta(\Lambda_2)]A_\mu{}^M = \delta(\Lambda_3)A_\mu{}^M - Y^M{}_{PQ}\left(\Lambda_1^{(P}{ D}_\mu\Lambda_2^{Q)}-\Lambda_2^{(P}{D}_\mu\Lambda_1^{Q)}
\right)\,,
\end{equation}
with
\begin{equation}\label{deflambdaLL}
 \Lambda_{3}^M \equiv X_{[PQ]}{}^M \Lambda_1^P \Lambda_2^Q\,.
\end{equation}
We see that, in order for the algebra to close on the gauge fields, they should have a transformation that is proportional to $Y^M{}_{PQ}$. This is exactly the $\delta(\Xi)$-transformation and
we find
\begin{equation}
 \label{commLLA} [\delta(\Lambda_1),\delta(\Lambda_2)]A_\mu{}^M = \delta(\Lambda_{3})A_\mu{}^M + \delta(\Xi_{3})A_\mu{}^M\,,
\end{equation}
with
\begin{equation}\label{defxiLL}
 \delta(\Xi_{3})A_\mu{}^M=-Y^M{}_{PQ} \Xi_{3\mu}{}^{PQ}\,,\qquad \Xi_{3\mu}{}^{PQ}\equiv \Lambda_1^{(P}{D}_\mu \Lambda_2^{Q)}-\Lambda_2^{(P}{D}_\mu\Lambda_1^{Q)}\,.
\end{equation}

The next step in the construction of a gauge invariant theory is the introduction of covariant field strengths for the vectors. As we mentioned before, the usual expression, (\ref{ordinaryF}), does not transform covariantly but picks up terms that are proportional to $Y^M{}_{NP}$. Therefore, we will introduce new field strengths,
\begin{equation}\label{covH}
 {\cal H}_{\mu\nu}{}^M\equiv {\cal F}_{\mu\nu}{}^M+Y^M{}_{NP}B_{\mu\nu}{}^{NP}\,,
\end{equation}
and new $2$-forms $B_{\mu\nu}{}^{NP}$.

Under the \textit{rigid symmetry group}, the $2$-forms transform in the symmetric product of two vector representations,
$(R_{\rm vect}\times R_{\rm vect})_{\rm symm}$. However, since they are contracted with a tensor $Y^M{}_{NP}$, part of the $2$-forms might be projected out. More precisely, if we introduce a projector\footnote{\label{fn:projector}The only property of ${\mathbb P}^{RS}{}_{NP}$ that we will use in this text is ${\mathbb P}^{RS}{}_{[NP]}=0$. So in principle the projector could be chosen as trivial ${\mathbb P}^{RS}{}_{NP}=\delta_{(N}{}^R\delta_{P)}{}^S$. Below, however, we will discuss constraints for the embedding tensor. If the latter satisfies the constraints, $Y^M{}_{NP}$ may have less components than all symmetric $(NP)$ combinations. ${\mathbb P}^{RS}{}_{NP}$ can then be chosen of lower rank such that it projects only to the $(NP)$ that remain in $Y^M{}_{NP}$.} ${\mathbb P}^{RS}{}_{NP}$ that leaves $Y^M{}_{NP}$ invariant,
\begin{equation}\label{defP2}
 Y^M{}_{RS}{\mathbb P}^{RS}{}_{NP}=Y^M{}_{NP}\,,
\end{equation}
then the only $2$-forms that survive are the ones that do not vanish upon contraction with ${\mathbb P}^{RS}{}_{NP}$.
We will use the special notation $B_{\mu\nu}{}^{\lc RS \rf}$ for these non-vanishing tensors:\footnote{This notation was first introduced in \cite{deWit:2008ta}.}
\begin{equation}\label{defPB}
  B_{\mu\nu}{}^{\lc RS \rf} \equiv {\mathbb P}^{RS}{}_{NP}B_{\mu\nu}{}^{NP}\,,\quad \mbox{ and }\quad Y^M{}_{NP}B_{\mu\nu}{}^{NP}=Y^M{}_{NP}B_{\mu\nu}{}^{\lc NP \rf}\,.
\end{equation}
The same reasoning applies to the parameters $\Xi_\mu{}^{MN}$, which can be restricted to $\Xi_\mu{}^{\lc MN\rf}$.

The \textit{gauge transformations} of the $B_{\mu\nu}{}^{\lc NP \rf}$ are fixed by demanding the covariant transformation of the new field strengths ${\cal H}_{\mu\nu}{}^M$.
More explicitly, we determine $\delta(\Lambda,\Xi) B_{\mu\nu}{}^{\lc NP  \rf}$ such that
\begin{equation}\label{deltaH}
 \delta(\Lambda,\Xi) {\cal H}_{\mu\nu}{}^M = - \Lambda^P X_{PN}{}^M  {\cal H}_{\mu\nu}{}^N\,.
\end{equation}
We find
\begin{equation}\label{deltaB}
 \delta(\Lambda,\Xi) B_{\mu\nu}{}^{NP} = 2 {D}_{[\mu}\Xi_{\nu]}{}^{NP} + 2 A_{[\mu}{}^{\lc N}\delta A_{\nu]}{}^{P \rf} - 2 \Lambda^{\lc N}{\cal H}_{\mu\nu}{}^{P \rf}\,,
\end{equation}
where we introduced the general notation
\begin{equation} \label{specialbrackets}
  A^{\lc M} B^{N \rf}\equiv {\mathbb P}^{MN}{}_{RS}A^{R} B^{S}\,,
\end{equation}
for some tensors $A^{R}$ and $B^{S}$. The transformations (\ref{deltaB}) can always be supplemented by extra terms that vanish upon contraction with $Y^M{}_{NP}$. Since the $2$-forms will be contracted with $Y^M{}_{NP}$ in the remainder of this and the next subsection, we will not consider these extra terms here. However, they will become important in section \ref{s:closure} where we compute the algebra on the $2$-forms, similar to our discussion for the vectors in (\ref{algebraLLA})-(\ref{defxiLL}).

\subsection{Gauge invariant action in $D=4$ \label{s:4daction}}
We have now introduced the minimal amount of ingredients that led to the construction of a consistent gauge algebra and covariant field strengths. In addition to the vector fields and the $\delta(\Lambda)$-transformations, consistency required the introduction of additional local transformations and new $2$-form tensor fields. So far, we have not introduced any dynamics for the fields, or made reference to an action. However, as we alluded to in the introduction, the discussion in the previous section
can be embedded into a framework where such an action is present. A general expression in $D=4$ was first given in \cite{deWit:2005ub} and contains kinetic and
Chern-Simons terms for the vectors, topological terms for the 2-forms and general matter terms. We will use a similar\footnote{In \cite{deWit:2005ub}, a different
basis for the $2$-forms was introduced. More precisely, for every rigid symmetry generator there is a corresponding tensor $B_{\mu\nu \, \alpha}$. Our action, which is taken from \cite{DeRydt:2008hw}, depends on the $2$-forms $B_{\mu\nu}{}^{\lc MN \rf}$.} expression here that is a
functional of the fields $A_{\mu}{}^M$ and $B_{\mu\nu}{}^{\lc MN \rf}$:
\begin{equation}\label{action4d}
 {\cal L}_0= {\cal L}_{{\rm g.k.}}+{\cal L}_{\rm
  GCS}+{\cal L}_{{\rm top},B}+{\cal L}_{\rm matter}\\
\end{equation}
with
\begin{eqnarray}
\nonumber\mathcal{L}_{{\rm
g.k.}}&=&\ft{1}{4}e\mathcal{I}_{\Lambda\Sigma}\mathcal{H}_{\mu\nu}{}^{\Lambda}\mathcal{H}^{\mu\nu\Sigma}
-\ft{1}{8}\mathcal{R}_{\Lambda\Sigma}\varepsilon^{\mu\nu\rho\sigma}\mathcal{H}_{\mu\nu}{}^{\Lambda}\mathcal{H}_{\rho\sigma}{}^{\Sigma}\,,\\
\nonumber{\cal L}_{\rm GCS} &=&\varepsilon^{\mu\nu\rho\sigma}A_{\mu}{}^{M} A_{\nu}{}^{N}\left(
\ft13\,  X_{MN\,\Lambda}\,\partial_{\rho} A_{\sigma}{}^{\Lambda}
+\ft16 X_{MN}{}^{\Lambda}\partial_{\rho} A_{\sigma}{}_{\Lambda}
+\ft18 X_{MN\,\Lambda}X_{PQ}{}^{\Lambda} A_{\rho}{}^{P}A_{\sigma}{}^{Q}\right)\,,\\
\nonumber {\cal L}_{{\rm top},B}&=& \ft{1}{4}
  \varepsilon^{\mu\nu\rho\sigma}\,
  Y^\Lambda{}_{NP} \,B_{\mu\nu}{}^{\lc NP\rf} \,
  \Big({\cal F}_{\rho \sigma \,\Lambda }
  +\ft{1}{2}\,Y_{\Lambda RS} \,B_{\rho \sigma }{}^{\lc RS\rf} \Big)\,.
\end{eqnarray}
The tensors $\mathcal{I}_{\Lambda\Sigma}$ and $\mathcal{R}_{\Lambda\Sigma}$ are the real and imaginary part of the scalar dependent gauge kinetic function.
They also appear in the expression for the dual fields strengths:
\begin{equation}
 {\cal G}_{\mu\nu}{}^M =({\cal H}_{\mu\nu}{}^\Lambda,{\cal G}_{\mu\nu\Lambda})\quad \mbox{with }{\mathcal G}_{\mu\nu\,\Lambda} \equiv \varepsilon_{\mu\nu\rho\sigma}\frac{\partial {\cal L}}{\partial {\cal H}_{\rho\sigma}{}^\Lambda} = {\cal R}_{\Lambda\Gamma} {\cal
  H}_{\mu\nu}{}^{\Gamma} +\frac{1}{2} e\,\varepsilon_{\mu\nu\rho\sigma}\,
  {\cal I}_{\Lambda\Gamma}\, {\cal H}{}^{\rho\sigma\,\Gamma}\,.
\end{equation}
The precise form of the Lagrangian (\ref{action4d}) will be important in section \ref{s:gaugealgebra}, once we need the equations of motion for the fields.
A short calculation shows that they have the following form:
\begin{eqnarray}
\label{eomA} \frac{\partial S_0}{\partial A_\mu{}^M}&=&\frac{1}{2} \varepsilon^{\mu\nu\rho\sigma}\Omega_{MN}D_\nu {\cal G}_{\rho\sigma}{}^N - j^\mu_M \approx 0\,,\\
\label{eomB}\frac{\partial S_0}{\partial B_{\mu\nu}{}^{\lc MN\rf}}&=&\frac{1}{4} \varepsilon^{\mu\nu\rho\sigma}\Omega_{RS}Y^R{}_{MN}\left({\cal H}-{\cal G}\right)_{\rho\sigma}{}^S \approx 0\,,
\end{eqnarray}
with $j^\mu_M \equiv \frac{\partial S_{\rm matter}}{\partial A_\mu{}^M}$ and identifications on-shell are indicated by $\approx$.

In \cite{DeRydt:2008hw} it was pointed out that the Lagrangian in (\ref{action4d}) is not automatically gauge invariant under the transformations (\ref{deltaA}) and
(\ref{deltaB}). Indeed, the structure that we studied in the previous section has to be supplemented by two extra ingredients.
\begin{enumerate} \item The embedding tensor has to satisfy a second constraint, known in the literature as the linear or representation constraint.
It has the following form:
\begin{equation}\label{linearconstr}
  D_{MNP}\equiv X_{(MN}{}^Q\Omega_{P)Q}=0\,.
\end{equation}
This constraint was first found as a necessary condition for supersymmetry invariance of theories with a maximal amount of supercharges. However, it also plays a crucial role in showing gauge invariance of the action (\ref{action4d}).

In \cite{DeRydt:2008hw}, a more precise meaning was attached to the linear constraint since it was recognized as the condition for the absence of quantum gauge anomalies. In the presence of gauge anomalies, characterized by constants $d_{\alpha\beta\gamma}$, this constraint should be modified to
\begin{equation}
D_{MNP}=\Theta_M{}^\alpha \Theta_N{}^\beta \Theta_P{}^\gamma d_{\alpha\beta\gamma}\,,
\end{equation}
and it leads to a Green-Schwarz cancellation mechanism. In this text, we will not consider anomalies and therefore, we use the form in (\ref{linearconstr}), i.e. $D_{MNP}=0$.

As an aside, we note that if we impose the linear constraint on the embedding tensor, the projectors ${\mathbb P}^{MN}{}_{RS}$ in (\ref{defP2}) take a special form. One can show that the corresponding set of 2-forms $\{B_{\mu\nu}{}^{\lc MN\rf}\}$ is given by $\{B_{\mu\nu\, {\widetilde \alpha}}\}$, where the $\{{\widetilde \alpha}\}$ are those indices for which $(t_{\widetilde \alpha})_M{}^N \neq 0$. In the literature (see e.g. \cite{deWit:2005ub}), the set $\{B_{\mu\nu\, {\widetilde \alpha}}\}$ is often extended to a basis with a 2-form for each adjoint index, i.e.
\begin{equation}
\left\{B_{\mu\nu\, {\widetilde \alpha}}\right\}  \quad \rightarrow \quad \left\{B_{\mu\nu\, {\alpha}}\right\}\,.
\end{equation}

 If one wants to write down a gauge invariant action for this extended set of 2-forms, a third constraint needs to be introduced on the embedding tensor. This constraint is called the locality constraint.

\item The second modification concerns the transformations of the $2$-forms. They need to be supplemented by extra terms that reflect the dependence of the Lagrangian on the matter content.
So once we specify the dynamics for the fields, the transformations of the $2$-forms are
\begin{equation}\label{fulldeltaB}
 \delta(\Lambda, \Xi) B_{\mu\nu}{}^{\lc NP\rf} = 2 {D}_{[\mu}\Xi_{\nu]}{}^{\lc NP\rf} + 2 A_{[\mu}{}^{\lc N}\delta A_{\nu]}{}^{P\rf} - 2 \Lambda^{\lc N}{\cal H}_{\mu\nu}{}^{P \rf}+\Delta B_{\mu\nu}{}^{\lc NP\rf}\,,
\end{equation}
with
\begin{equation}\label{DeltaB}
 \Delta B_{\mu\nu}{}^{\lc NP\rf}=- 2 \Lambda^{\lc N}\left({\cal G}_{\mu\nu}{}^{P \rf}-{\cal H}_{\mu\nu}{}^{P \rf}\right)
\end{equation}
which is a scalar field dependent quantity. Again, (\ref{fulldeltaB}) can be extended by adding extra terms that vanish upon contraction with $Y^M{}_{NP}$.
\end{enumerate}
To summarize, the action (\ref{action4d}) is invariant under the new set of gauge transformations (\ref{deltaA}) and (\ref{fulldeltaB}),
provided that we use the closure and linear constraints on the embedding tensor.

Let us also repeat here that if we do not stipulate an action, the transformations (\ref{deltaB}) are perfectly feasible and also lead to a consistent gauge algebra. In the future, we will always consider both cases for $\Delta B_{\mu\nu}{}^{\lc NP \rf}$: it is either zero as in (\ref{deltaB}), or given by (\ref{DeltaB}).

\section{Structure of the gauge algebra \label{s:gaugealgebra}}
We are now ready to give more details on the gauge algebra structure that arises in the
$D=4$ embedding tensor formalism. We will provide an answer to the following questions:
\begin{enumerate}
 \item Is the algebra closed? For a closed algebra, the commutator of two gauge transformations leads again to a linear combination of transformations, with new parameters that depend on the fields and the original parameters. Part of this question was already answered in section \ref{s:structure}, where we checked that the algebra closes on the vectors with transformations $\delta(\Lambda)$ and $\delta(\Xi)$. Here we will extend this result to the 2-forms and show that the algebra only closes in the absence of an action (i.e. $\Delta B = 0$). In the other case (i.e. $\Delta B \neq 0$) we encounter an open algebra, where additional terms in the commutator appear that are proportional to the field equations.
\item Are the structure constants really `constant', or are they functions of the fields? The gauge algebra in the embedding tensor formalism turns out to be a soft algebra.
 \item Is the gauge algebra (ir)reducible? This question addresses the (in)dependence of the different gauge transformations, which is important if we want to determine the independent degrees of freedom in the theory. We show that our algebra is higher stage reducible.
\end{enumerate}
These issues will be dealt with in several steps. In section \ref{s:dewitt}, we first introduce some useful notation, and discuss the gauge algebra generators. In the next section \ref{s:closure}, we complete the discussion on the gauge algebra commutators and show that the algebra is closed for $\Delta B = 0$ and open for $\Delta B \neq 0$. Once all the commutators are known, the structure `constants' can easily be determined. They will turn out to be field dependent and thus lead to a soft algebra. Finally, in \ref{s:zeromodes} and \ref{s:higherstagezerom} we investigate the dependencies of the gauge transformations and construct the zero modes for the reducible algebra.

\subsection{DeWitt notation and gauge generators \label{s:dewitt}}
In order to facilitate our discussion, we will introduce the DeWitt notation that provides a compact and transparent way of writing down general field theories. The different fields are denoted by $\phi^i$, where the index $i=1,\ldots,n$ can label spacetime indices $\mu,\,\nu,\,\hdots$ for tensor fields, spinor indices for fermion fields, and/or an index distinguishing different types of generic fields. The fields are also functions of spacetime, and we will adopt the convention that the appearance of a discrete DeWitt index also indicates the presence of a spacetime variable. We then use a generalized summation convention in which a repeated discrete index implies not only a sum over that index but also an integration over the corresponding spacetime variable.

In our case, the general notation $\phi^i$ is an abbreviation for the collection of bosonic tensor fields,
\begin{equation}
 \phi^i\,\in\,\left\{A_\mu{}^M(x),\,B_{\mu\nu}{}^{\lc MN\rf}(x)\right\}\,.
\end{equation}
It means that the $i$-index takes the following discrete values: $\left\{\mu M,\,{\mu\nu}{\lc MN\rf}\right\}$, where one should remember that the spacetime indices and (combinations of) vector indices are not at the same level.

Furthermore, we have a set of $m_0$ non-trivial bosonic gauge transformations. In the DeWitt notation, they take the following form \footnote{Without the use of a compact summation convention, this relation would be represented as
\begin{equation}
 \delta \phi^i(x)=\int {\rm d}y \,\R^i{}_{a_0}(x,y)\,\varepsilon^{a_0}(y)\,.
\end{equation}
}
\begin{equation}\label{generalgaugetransfo}
 \delta \phi^i=\R^i{}_{a_0}(\phi)\varepsilon^{a_0}\,,\qquad \mbox{with } {a_0}=1,\,2,\, \ldots,\,m_0\,.
\end{equation}
The infinitesimal gauge parameters $\varepsilon^{a_0}$ are arbitrary functions of the spacetime variable $x$, and $\R^i{}_{a_0}$ denotes the generators of the gauge transformations.
The different types of gauge parameters that we have introduced so far are:
\begin{equation}\label{parameters}
 \varepsilon^{a_0} \,\in\, \left\{\Lambda^M(x),\,\Xi_{\mu}{}^{\lc MN\rf}(x)\right\}\end{equation}
thus, ${a_0} \,\in\, \left\{M,\,\mu{\lc MN \rf}\right\}$.
The gauge generators $\R^i{}_{a_0}$ in the embedding tensor formalism can be computed by comparing the field transformations to the general expression (\ref{generalgaugetransfo}). For the vector fields, we find from (\ref{deltaA}) that their transformations are generated by \footnote{In the following, we will use the notation $D_{\mu}{}^{N_1\hdots N_p}{}_{M_1 \hdots M_p}$, which is a particular derivation operation. For $p=1$, it is defined as follows:
\begin{equation}
 D_\mu{}^N{}_M T^M \equiv \left(\delta_M{}^N\partial_\mu + A_\mu{}^Q X_{Q M}{}^N\right) T^M = D_\mu T^N\,,
\end{equation}
for some object $T^M$ that transforms with a vector index. A general definition for arbitrary $p$ can be found in (\ref{specder}), together with some of its properties.}
\begin{eqnarray}
\label{genALambda} \R_{\mu}{}^M{}_K&=&D_{\mu}{}^M{}_K  \label{GG1}\,, \\
\label{genAXi} \R_{\mu}{}^{M\,\nu}{}_{\lc KL\rf}&=&-\delta^{\nu}_{\mu}\,Y^{M}{}_{KL}\,.  \label{GG2}
\end{eqnarray}
The $\mu M$-indices in these equations correspond to the $i$-index in $\R^i{}_{a_0}$. The  $a_0$-index takes the values $K$ in (\ref{genALambda}) and $\nu \lc KL\rf$ in (\ref{genAXi}). Let us for once give a more detailed discussion on how (\ref{GG1}) and (\ref{GG2}) are obtained (all the subsequent results can be found in a similar way). We choose $\phi^i$ to be equal to a vector, then (\ref{generalgaugetransfo}) can be written as
\begin{eqnarray}
  \delta A_\mu{}^M(x)&=&\int {\rm d}^4y\, \R_\mu{}^M{}_{a_0}(x,y) \varepsilon^{a_0}(y)\\
  &=&\int {\rm d}^4y \left[\R_\mu{}^M{}_K(x,y) \Lambda^{K}(y)+\R_\mu{}^{M\,\nu}{}_{\lc KL\rf}(x,y)\Xi_\nu{}^{\lc KL\rf}(y) \right]
\end{eqnarray}
In the first line, we explicitly wrote down the integral over the spacetime variable $y$, which was hidden in the summation over $a_0$. In the second line, we further worked out the summation over $a_0$. This result should now be compared to (\ref{deltaA}), and we find
\begin{eqnarray}
  \nonumber \R_\mu{}^M{}_K(x,y) &=& \left(\delta_K{}^M \frac{\partial}{\partial x^\mu}-A_\mu{}^Q(x)X_{QK}{}^M\right)\delta(x-y)\\&=& \label{calcRA1} D_\mu{}^M{}_K(x)\delta(x-y)\,,\\
  \label{calcRA2} \R_\mu{}^{M\,\nu}{}_{\lc KL\rf}(x,y)&=&-\delta_\mu^\nu Y^M{}_{KL}\delta(x-y)\,.
\end{eqnarray}
In the following, we will suppress the spacetime variables and delta functions, such that (\ref{calcRA1}) and (\ref{calcRA2}) reduce to the expressions (\ref{genALambda}) and (\ref{genAXi}).

Likewise, the tensors that generate the transformations of the 2-forms $B_{\mu \nu}{}^{\lc NP\rf}$ can be determined from (\ref{fulldeltaB}):
\begin{equation}
  \begin{array}{rcll}
\vspace{2mm}\label{Bgenerators}\multirow{2}{*}{$\left\{\begin{array}{rr}\vspace{1mm}\R_{\mu \nu}{}^{  \lc NP \rf }{}_K\\ \RS_{\mu \nu}{}^{ \lc NP \rf}{}_K\end{array}\right.$}&=&2A_{[\mu}{}^{ \lc N}D_{\nu]}{}^{P \rf }{}_K-2\delta_K{}^{\lc N}\,{\cal H}_{\mu\nu}{}^{ P \rf}&\hspace{-5mm}\mbox{ for}\;\; \Delta B_{\mu\nu}{}^{\lc NP\rf} =0\,,\\\vspace{2mm}&=&2A_{[\mu}{}^{ \lc N}D_{\nu]}{}^{ P \rf }{}_K-2\delta_K{}^{\lc N} \,{\cal G}_{\mu\nu }{}^{P \rf}&\hspace{-5mm}\mbox{ for}\;\; \Delta B_{\mu\nu}{}^{\lc NP\rf} =- 2 \Lambda^{\lc N}\left({\cal G}-{\cal H}\right)_{\mu\nu}{}^{P \rf}, \\\vspace{2mm}
\R_{\mu \nu}{}^{\lc NP \rf}{}^{\rho}{}_{ \lc RS \rf}&=&2D_{[\mu}{}^{ NP}{}_{RS}\delta_{\nu]}^{\rho}-2A_{[\mu}{}^{\lc N}Y^{P \rf}{}_{RS}\delta_{\nu]}^{\rho}\,.&
\end{array}
\end{equation}
In the first two lines, we made a distinction between the gauge transformations without the specification of an action (i.e. $\Delta B=0$), and the gauge transformations that leave the action (\ref{action4d}) invariant. In order to tell the difference between these two cases, we have added a tilde to the generators on the second line. In the following, we will use the more general notation $\RS{}^i{}_{a_0}$ to indicate all generators that leave the action (\ref{action4d}) invariant. It is clear that $\RS{}^i{}_{a_0}={\R}{}^i{}_{a_0}$, except for $i=\mu \nu \lc NP \rf$, $a_0=K$.

Given the precise form of the gauge generators, the next step is to compute the commutators of gauge transformations on the fields. This will be the subject of the next section.

\subsection{Closure of the gauge algebra \label{s:closure}}
We make a clear distinction between the generators $\R{}^i{}_{a_0}$ that are part of the embedding tensor formalism without the specification of an action, and the generators $\RS{}^i{}_{a_0}$ that appear in the Lagrangian description. The difference between these two cases, which is captured by $\Delta B$, leads to distinct conclusions about the properties of the corresponding gauge algebra. In the first subsection, we consider the formalism without an action.

\subsubsection{Formalism without an action}
In order to have a closed algebra, the gauge transformations need to satisfy the following relation:
\begin{equation}\label{closure}
 [\delta_1,\delta_2]\phi^i = \R{}^i{}_{a_0} \T^{a_0}{}_{b_0 c_0}\,\varepsilon_1^{b_0}\varepsilon_2^{c_0}\,.
\end{equation}
The $\T^{a_0}{}_{b_0 c_0}$ are antisymmetric tensors under the interchange of indices $b_0$ and $c_0$. They are called the `structure constants' of the algebra, although in general, they depend on the fields of the theory.

From (\ref{commLLA}) we know that the commutator of two $\delta(\Lambda)$ transformations on the gauge fields leads again to a linear combination of a $\delta(\Lambda)$ and a $\delta(\Xi)$ transformation. Likewise, one can show that
\begin{eqnarray}
  \label{commLXA}[\delta(\Lambda),\delta(\Xi)]A_\mu{}^M&=&0\,,\\
 \label{commXXA}[\delta(\Xi_1),\delta(\Xi_2)]A_\mu{}^M&=&0\,.
\end{eqnarray}
We conclude that the gauge algebra with transformations $\delta(\Lambda)$ and $\delta(\Xi)$ indeed satisfies the relation (\ref{closure}) on the vector fields.
The only non-vanishing structure constants are
\begin{equation}\label{strfunctionA}
 \T^M{}_{RS}=X_{[RS]}{}^M\,,\qquad
\T_\mu{}^{\lc MN\rf}{}_{RS}=\delta_R{}^{\lc M}D_{\mu}{}^{N\rf}{}_{S}-\delta_S{}^{\lc M}D_{\mu}{}^{N\rf}{}_{R}\,,
\end{equation}
where both $\T$'s are antisymmetric in $[RS]$.

On the other hand, we have not yet checked whether the algebra also closes on the 2-forms. Let us therefore compute the non-vanishing commutators. We find that:
\begin{eqnarray}
 \label{commLLB} [\delta(\Lambda_1),\delta(\Lambda_2)]B_{\mu\nu}{}^{\lc NP\rf}&=& \delta(\Lambda_{3})B_{\mu\nu}{}^{\lc NP\rf} + \delta(\Xi_{3})B_{\mu\nu}{}^{\lc NP\rf}\\
\nonumber&&-Y^{NP}{}_{M\lc RS\rf}\left(\Lambda_1^M {\cal H}_{\mu\nu}{}^{\lc R}\Lambda_2^{S\rf}
-(1\leftrightarrow 2)\right)\,,\\
 \label{commLXB} [\delta(\Lambda),\delta(\Xi)]B_{\mu\nu}{}^{\lc NP\rf}&=&-Y^{NP}{}_{M\lc RS\rf}\left(-2 \Xi_{[\mu}{}^{\lc RS\rf}{ D}_{\nu]}\Lambda^M\right)\,,\\
 \label{commXXB} [\delta(\Xi_1),\delta(\Xi_2)]B_{\mu\nu}{}^{\lc NP\rf}&=&-Y^{NP}{}_{M\lc RS\rf}\left[-Y^M{}_{QT}\left(\Xi_{1[\mu}{}^{\lc QT\rf}\Xi_{2\nu]}{}^{\lc RS\rf}-(1\leftrightarrow2)\right)\right]\,,
\end{eqnarray}
with
\begin{equation}\label{defY2}
 Y^{NP}{}_{M\lc RS\rf} \equiv 2\left(\delta_M{}^{\lc N}Y^{P \rf}{}_{RS}-X_{M \lc R}{}^{\lc N}\delta_{S\rf}{}^{P\rf}\right).
\end{equation}
The contraction of this tensor with $Y^Q{}_{NP}$ vanishes (see (\ref{YY1})),
\begin{equation}\label{contrY2Y3}
 Y^Q{}_{NP} Y^{NP}{}_{M\lc RS\rf} = 0\,,
\end{equation}
which is a relation that will be important later on.

Let us now study the commutation relations (\ref{commLLB})-(\ref{commXXB}) in more detail. Clearly, the closure condition (\ref{closure}) is not satisfied since each of the commutators contains an extra term that is proportional to $Y^{NP}{}_{M\lc RS\rf}$. There is however a way to restore closure of the algebra, which is completely analogous to our treatment in (\ref{algebraLLA})-(\ref{defxiLL}) for the 1-forms: we extend the original gauge transformations
with a new local transformation that is proportional to $Y^{NP}{}_{M\lc RS\rf}$,
\begin{equation}\label{newdeltaB}
 \delta(\Lambda,\Xi)B_{\mu\nu}{}^{ \lc MN\rf}\quad \rightarrow \quad\delta(\Lambda,\Xi,\Phi)B_{\mu\nu}{}^{\lc MN\rf}=\delta(\Lambda,\Xi)B_{\mu\nu}{}^{\lc MN\rf}+\delta(\Phi)B_{\mu\nu}{}^{\lc MN\rf},
\end{equation}
with \footnote{We have used a special notation here with nested brackets $\lc.\lc.\rf\rf\,$. One can think of it as a generalization of $\lc.\rf\,$; its precise definition and further generalizations can be found in appendix \ref{a:covcontra}.}
\begin{equation}
\delta(\Phi)B_{\mu\nu}{}^{\lc MN\rf}=-Y^{MN}{}_{P\lc RS\rf}\Phi_{\mu\nu}{}^{\lc P \lc RS\rf\rf}
\end{equation}
and new local parameters $\Phi_{\mu\nu}{}^{\lc P\lc RS\rf\rf}(x)$.
The original set of gauge transformations in (\ref{parameters}) should therefore be replaced by $\{\delta(\Lambda),\delta(\Xi),\delta(\Phi)\}$ with local parameters
\begin{equation}\label{extendeda0}
 \varepsilon^{a_0} \,\in\, \left\{\Lambda^M(x),\,\Xi_{\mu}{}^{\lc MN\rf}(x),\,\Phi_{\mu\nu}{}^{\lc M \lc NP\rf\rf}(x)\right\}\,.
\end{equation}
It is clear that the index $a_0$ takes an extra value $\mu\nu \lc M \lc NP\rf\rf$ and the corresponding gauge generators are
\begin{eqnarray}
\R_{\rho}{}^{K}{}^{\,\mu\nu}{}_{  \lc M \lc NP\rf\rf }&=&0\,,\\
 \R_{\rho\sigma}{}^{ \lc KL \rf \,\mu\nu}{}_{ \lc M  \lc NP\rf \rf}&=&-\delta^{[\mu\nu]}_{\rho \sigma}Y^{KL}{}_{M  \lc NP \rf}\,.
\end{eqnarray}
With this new set of gauge transformations, it is easy to check that the algebra closes:
\begin{eqnarray}
\label{commLLB2}[\delta(\Lambda_1),\delta(\Lambda_2)]B_{\mu\nu}{}^{\lc NP\rf}&=& \delta(\Lambda_{3})B_{\mu\nu}{}^{\lc NP\rf} + \delta(\Xi_{3})B_{\mu\nu}{}^{\lc NP\rf}+\delta(\Phi_{3})B_{\mu\nu}{}^{\lc NP\rf}
\,,\\
\label{commLXB2}[\delta(\Lambda),\delta(\Xi)]B_{\mu\nu}{}^{\lc NP\rf}&=&\delta(\Phi_{4})B_{\mu\nu}{}^{\lc NP\rf}\,,\\
\label{commXXB2} [\delta(\Xi_1),\delta(\Xi_2)]B_{\mu\nu}{}^{\lc NP\rf}&=&\delta(\Phi_{5})B_{\mu\nu}{}^{\lc NP\rf}\,.
\end{eqnarray}
The commutators satisfy the relation (\ref{closure}) and the parameters $\Phi_{3}$, $\Phi_{4}$ and $\Phi_{5}$ determine the precise form of the structure constants. From (\ref{commLLB}), (\ref{commLXB}) and (\ref{commXXB}) one sees that:
\begin{eqnarray}
  \Phi_{3}{}_{\mu\nu}{}^{\lc M\lc RS\rf\rf}&=&\Lambda_1^{\lc M} {\cal H}_{\mu\nu}{}^{\lc R}\Lambda_2^{S\rf\rf}-(1\leftrightarrow 2)\,,\\
  \Phi_{4}{}_{\mu\nu}{}^{\lc M\lc RS\rf\rf}&=&-2 \Xi_{[\mu}{}^{\lc\lc RS\rf}{ D}_{\nu]}\Lambda^{M\rf}\,,\\
  \Phi_{5}{}_{\mu\nu}{}^{\lc M\lc RS\rf\rf}&=&Y^{\lc M}{}_{QT}\left(\Xi_{1[\mu}{}^{\lc RS\rf\rf}\Xi_{2\nu]}{}^{QT}-(1\leftrightarrow2)\right)\,,
\end{eqnarray}
which leads to an expression for the remaining non-vanishing structure constants:
\begin{eqnarray}
\label{strfunctionB1} \T_{\mu\nu}{}^{\lc M\lc NP\rf\rf}{}_{RS}&=&\delta_R{}^{\lc M} {\cal H}_{\mu\nu}{}^{\lc N}\delta_S{}^{P\rf\rf} -\delta_S{}^{\lc M} {\cal H}_{\mu\nu}{}^{\lc N}\delta_R{}^{P\rf\rf} \,,\\\label{strfunctionB2}
\T_{\mu\nu}{}^{\lc M\lc NP\rf\rf}{}_{R}{}^{\,\rho}{}_{\lc ST\rf}&=&-\T_{\mu\nu}{}^{\lc M\lc NP\rf\rf\,\rho}{}_{\lc ST\rf\, R}\;=\;-2\delta_S{}^{\lc\lc N} \delta_T{}^{P\rf} \delta_{[\mu}^\rho D_{\nu]}{}^{M\rf}{}_R\,,\\\label{strfunctionB3}
\T_{\mu\nu}{}^{\lc M\lc NP\rf\rf\,\rho}{}_{\lc QR\rf}{}^\sigma{}_{\lc ST\rf}&=&\delta_{[\mu}^\rho \delta_{\nu]}^\sigma \left(Y^{\lc M}{}_{QR}\delta_S{}^{\lc N}\delta_T{}^{P\rf\rf}+Y^{\lc M}{}_{ST}\delta_Q{}^{\lc N}\delta_R{}^{P\rf\rf}\right)\,.
\end{eqnarray}

We conclude this section with the observation that the structure constants are not really constant, but depend on the fields in the theory. This can be seen for example from (\ref{strfunctionA}) which depends on the vectors, and (\ref{strfunctionB1}) which depends on both the vectors and $2$-forms. Due to this field dependence of the structure functions, the gauge algebra is often called a `soft algebra'.

\subsubsection{Formalism with an action}
Our treatment in the previous section can now be generalized to the case where the embedding tensor formalism is incorporated into the framework of an action. This has an effect on the gauge transformations, i.e. $\Delta B_{\mu\nu}{}^{\lc MN\rf}$ takes the value in (\ref{DeltaB}) and the generators $\R{}^i{}_{a_0}$ are replaced by $\RS{}^i{}_{a_0}$. So far, the latter have only been defined for the indices $i \in \{\mu M,\,\mu\nu \lc MN\rf\}$ and $a_0 \in\{M, \mu \lc MN\rf\}$. We will see in due course that the $a_0$ have to be extended as in (\ref{extendeda0}), but for the time being, we only consider the smaller set. In section \ref{s:4daction} we saw that the transformations $\RS{}^i{}_{a_0}$ leave the action invariant, which is expressed by the Noether identities:
\begin{equation}\label{Noether}
 \partial_i S_{0}\,\RS{}^i{}_{a_0} = 0\,.
\end{equation}
The most general solution to the Noether identities is a gauge transformation, up to terms proportional to the equations of motion:
\begin{equation}\label{solNoether}
 \partial_i S_{0}\,\lambda^i = 0 \quad\Leftrightarrow\quad \lambda^i = \RS{}^i{}_{a_0} {\chi}^{a_0} + \partial_j S_{0}\, T^{ji}\,,
\end{equation}
for some tensors ${\chi}^{a_0}$ and $T^{ij}=-T^{ji}$. The last term in (\ref{solNoether}) is known as a trivial gauge transformation, and it is easily checked that the action is invariant under these transformations due to the antisymmetry of $T^{ij}$ in $[ij]$. A particular choice for $\lambda^i$ in (\ref{solNoether}) would be the commutator of two gauge transformations on a field: $\lambda^i = [\delta_1,\delta_2]\phi^i$. Since for this particular choice, $\partial_i S_{0}\, \lambda^i=0$ is trivially satisfied due to (\ref{Noether}), equation (\ref{solNoether}) tells us that  $[\delta_1,\delta_2]\phi^i$ is of the form
\begin{equation}
[\delta_1,\delta_2]\phi^i = \RS{}^i{}_{a_0} {\chi}^{a_0} + \partial_j S_{0}\, T^{ji}\,.
\end{equation}
Since the left hand side is proportional to the antisymmetric combination of two gauge parameters, $\varepsilon_1^{[a_0}\varepsilon_2^{b_0]}$, so should be the right hand side. We can factor out these parameters and write
\begin{equation}\label{offshellalgebra}
[\delta_1,\delta_2]\phi^i = \RS{}^i{}_{a_0} \TS{}^{a_0}{}_{b_0 c_0}\,\varepsilon_1^{b_0}\varepsilon_2^{c_0} + \partial_j S_{0}\, \ES{}^{ij}{}_{a_0 b_0}\,\varepsilon_1^{a_0}\varepsilon_2^{b_0}\,.
\end{equation}
This is the generalization of equation (\ref{closure}). The first term on the right hand side has a familiar form, with $\TS{}^{a_0}{}_{b_0 c_0}$ the structure `constants' that are antisymmetric in $[b_0c_0]$. The second term depends on the equations of motion multiplied by some $\ES$-tensors that are antisymmetric in both $[ij]$ and $[a_0 b_0]$. If these tensors do not vanish, the algebra only closes on-shell (i.e. when $\partial_j S_{0} = 0$).

To summarize, the Noether identities impose a particular form for the gauge algebra, given by (\ref{offshellalgebra}). We will now check whether (\ref{offshellalgebra}) is indeed fulfilled for the 1- and 2-forms that arise in the embedding tensor formalism. Our results are as follows:

\paragraph{Commutators on the $1$-forms.} These do not change, i.e. (\ref{commLLA}), (\ref{commLXA}) and (\ref{commXXA}) are still valid. The reason is that the generators $\RS{}^i{}_{a_0}$ are equal to the generators $\R{}^i{}_{a_0}$ for $i=M$ and arbitrary $a_0$. This also means that the structure functions take the same values:
      \begin{equation}
       \TS{}^M{}_{RS}=\T^M{}_{RS}\,,\qquad
\TS{}_\mu{}^{\lc MN\rf}{}_{RS}=\T{}_\mu{}^{\lc MN\rf}{}_{RS}\,.
      \end{equation}
      The corresponding $\ES$-tensors in (\ref{offshellalgebra}) all vanish.

\paragraph{Commutators on the $2$-forms.} These are slightly more involved. Let us start with the easiest case, which is the commutator of a $\delta(\Lambda)$ and a $\delta(\Xi)$ transformation. We find
    \begin{equation}
      \label{offshellcommLXB}[\delta(\Lambda),\delta(\Xi)]B_{\mu\nu}{}^{\lc MN\rf}=-Y^{MN}{}_{Q\lc RS\rf}\left(-2 \Xi_{[\mu}{}^{\lc RS\rf}{ D}_{\nu]}\Lambda^Q\right)\,,
    \end{equation}
    which is exactly the same expression as (\ref{commLXB}). Consistency of the algebra requires the introduction of a new local transformation of the $2$-forms, identical to (\ref{newdeltaB}). The corresponding generators are
    \begin{eqnarray}
\label{generatorphi}\RS{}_{\mu}{}^{M\,\rho \sigma}{}_{  \lc Q \lc RS\rf\rf }&=&0\,,\\\label{generatorphi2}
 \RS{}_{\mu \nu}{}^{\lc  MN \rf\,\rho \sigma}{}_{ \lc Q  \lc RS\rf\rf }&=&-\delta^{[\rho \sigma]}_{\mu\nu}Y^{MN}{}_{Q  \lc RS \rf}\,,
\end{eqnarray} and the index range $a_0$ has to be extended to
\begin{equation}
  a_0 \in \left\{M, \mu \lc MN\rf,\,\mu\nu \lc M \lc NP\rf\rf\right\}\,.
\end{equation}

It is important to note that the form of the algebra in (\ref{offshellalgebra}) should still be valid for this extended set of indices. Since (\ref{offshellalgebra}) is a consequence of the Noether identities (\ref{Noether}), it is enough to check that the latter also hold for $a_0 = \mu\nu \lc M\lc NP\rf\rf$. Indeed,
\begin{eqnarray}
  \nonumber \partial_i S_{0}\,\RS{}^{i\,\mu\nu}{}_{ \lc M \lc NP\rf\rf} &=&\frac{\partial S_0}{\partial A_\rho{}^R} \,\RS{}_{\rho}{}^{R\,\mu\nu}{}_{\lc  M \lc NP\rf\rf }+\frac{\partial S_0}{\partial B_{\rho\sigma}{}^{\lc RS \rf}} \, \RS{}_{\rho\sigma}{}^{ \lc RS \rf\,\mu\nu}{}_{  \lc M  \lc NP \rf\rf }\\
  &\sim& \left(Y^Q{}_{RS}\right)\left(-\delta^{[\mu\nu]}_{\rho \sigma}Y^{RS}{}_{M  \lc NP \rf}\right)\quad = \quad 0\,.
\end{eqnarray}
We used (\ref{generatorphi}) and the fact that each $2$-form $B_{\mu\nu}{}^{\lc RS \rf}$ is contracted with a tensor $Y^Q{}_{RS}$ in the action. Then the last line vanishes because of the orthogonality of the $Y$-tensors, see (\ref{contrY2Y3}).

Due to the introduction of the new transformations $\delta(\Phi)$, relation (\ref{offshellcommLXB}) can be written as
\begin{equation}
  \label{offshellcommLXB2}[\delta(\Lambda),\delta(\Xi)]B_{\mu\nu}{}^{\lc MN\rf}=\delta({\widetilde \Phi}_{4})B_{\mu\nu}{}^{\lc MN\rf}\,,
\end{equation}
with ${\widetilde \Phi}_{4}=\Phi_{4}$. If we compare this to the general expression (\ref{offshellalgebra}), it is clear that all the corresponding $\ES$-tensors vanish and the non-zero structure function is identical to (\ref{strfunctionB2}):
\begin{equation}
 \TS{}_{\mu\nu}{}^{\lc M\lc NP\rf\rf}{}_{R}{}^\rho{}_{\lc ST\rf}=\T{}_{\mu\nu}{}^{\lc M\lc NP\rf\rf}{}_{R}{}^\rho{}_{\lc ST\rf}\,.
\end{equation}
Similar results hold for the commutator of two $\delta(\Xi)$ transformations on the $2$-forms. We have
\begin{equation}
  \label{offshellcommXXB2} [\delta(\Xi_1),\delta(\Xi_2)]B_{\mu\nu}{}^{\lc MN\rf}=\delta({\widetilde\Phi}_{5})B_{\mu\nu}{}^{\lc MN\rf}\,,
\end{equation}
with ${\widetilde \Phi}_{5}=\Phi_{5}$. Also here the $\ES$-tensors vanish and $\TS{}_{\mu\nu}{}^{\lc M\lc NP\rf\rf\,\rho}{}_{\lc QR\rf}{}^\sigma{}_{\lc ST\rf}$ is given by (\ref{strfunctionB3}).

Finally, we calculate the commutator of two $\delta(\Lambda)$ transformations:
\begin{eqnarray}
\nonumber [\delta(\Lambda_1),\delta(\Lambda_2)]B_{\mu\nu}{}^{\lc MN\rf}&=&\delta({\widetilde \Lambda}_{3})B_{\mu\nu}{}^{\lc MN\rf}
+\delta({\widetilde \Xi}_{3})B_{\mu\nu}{}^{\lc MN\rf}+\delta({\widetilde \Phi}_{3})B_{\mu\nu}{}^{\lc MN\rf}\\
\label{offshellcommLLB} &&+2\Lambda_1^{[Q}\Lambda_2^{P]}\Big{[}8\eta_{\rho[\mu}\eta_{\nu]\sigma} {\cal I}_{\Lambda\Sigma}\,{\mathbb P}^{MN}{}_P{}^\Lambda{\mathbb P}^{RS}{}_Q{}^\Sigma\\
\nonumber &&-4\varepsilon_{\mu\nu\rho\sigma}{\cal R}_{\Lambda\Sigma}\,{\mathbb P}^{MN}{}_P{}^\Lambda{\mathbb P}^{RS}{}_Q{}^\Sigma\\
 \nonumber&&-\varepsilon_{\mu\nu\rho\sigma}\left({\mathbb P}^{MN}{}_P{}^\Lambda{\mathbb P}^{RS}{}_{Q\Lambda}+{\mathbb P}^{MN}{}_{P\Lambda}{\mathbb P}^{RS}{}_Q{}^\Lambda\right)\Big{]}\frac{\partial S_0}{\partial B_{\rho\sigma}{}^{\lc RS\rf}}\,.
\end{eqnarray}
The parameters ${\widetilde \Lambda}_{3}$, ${\widetilde \Xi}_{3}$ and ${\widetilde \Phi}_{3}$ take the following values:
\begin{equation}
  {\widetilde \Lambda}_{3}^M=\Lambda_{3}^M\,,\quad {\widetilde \Xi}_{3}{}_\mu{}^{\lc MN\rf}=\Xi_{3}{}_\mu{}^{\lc MN\rf}\,,\quad {\widetilde \Phi}_{3}{}_{\mu\nu}{}^{\lc M\lc NP\rf\rf}=\Lambda_1^{\lc M} {\cal G}_{\mu\nu}{}^{ \lc N}\Lambda_2^{P\rf\rf}-(1\leftrightarrow 2)\,.
\end{equation}
This result is slightly more complicated and we note the following differences with (\ref{commLLB2}):
\begin{itemize}
 \item The parameter ${\widetilde \Phi}_{3}$ differs from $\Phi_{3}$, i.e. the field strengths ${\cal H}_{\mu\nu}{}^{M}$ have been replaced by their scalar dependent counterparts ${\cal G}_{\mu\nu}{}^{M}$.
\item This in turn leads to a difference in the structure functions:
\begin{equation}
 \TS{}_{\mu\nu}{}^{M\lc NP\rf}{}_{RS}=\delta_R{}^{\lc M} {\cal G}_{\mu\nu}{}^{\lc N}\delta_S{}^{P\rf\rf} -\delta_S{}^{\lc M} {\cal G}_{\mu\nu}{}^{\lc N}\delta_R{}^{P\rf\rf}\,.
\end{equation}
Again ${\cal H}_{\mu\nu}{}^{M}$ has been replace by ${\cal G}_{\mu\nu}{}^{M}$.
\item Finally, the last three lines in equation (\ref{offshellcommLLB}) are proportional to the equations of motion. These terms fit into the general expression (\ref{offshellalgebra}) with $\ES{}_{\mu\nu}{}^{MN}{}_{\rho\sigma}{}^{RS}{}_{PQ}$ different from zero:
\begin{eqnarray}
\nonumber \ES{}_{\mu\nu}{}^{MN}{}_{\rho\sigma}{}^{RS}{}_{PQ} &=&-16\eta_{\rho[\mu}\eta_{\nu]\sigma} {\cal I}_{\Lambda\Sigma}\,{\mathbb P}^{MN}{}_{[P}{}^\Lambda{\mathbb P}^{RS}{}_{Q]}{}^\Sigma
 +8\varepsilon_{\mu\nu\rho\sigma}{\cal R}_{\Lambda\Sigma}\,{\mathbb P}^{MN}{}_{[P}{}^\Lambda{\mathbb P}^{RS}{}_{Q]}{}^\Sigma\\
&&+2\varepsilon_{\mu\nu\rho\sigma}\left({\mathbb P}^{MN}{}_{[P}{}^\Lambda{\mathbb P}^{RS}{}_{Q]\Lambda}+{\mathbb P}^{MN}{}_{[P\Lambda}{\mathbb P}^{RS}{}_{Q]}{}^\Lambda\right)\,.
\end{eqnarray}
\end{itemize}
To summarize, let us repeat the main points of this section. If the embedding tensor formalism is modified by introducing a Lagrangian, we are necessarily dealing with an open algebra. The general form of such an algebra is given in (\ref{offshellalgebra}), and we checked that this relation is indeed satisfied. At the same time, the calculations provided us with an expression for the structure constants and $\ES$-tensors. This lets us conclude that we are also dealing with a soft algebra, i.e. the structure constants depend on the fields in the theory.

Of course this is not the end of the story. Several higher order commutators need to be evaluated in order to define the full structure of the algebra. For example, at the second order we find the Jacobi identity,
\begin{equation}\label{Jacobi}
[\delta_1,[\delta_2,\delta_3]]\phi^i+\mbox{cyclic in 123}=0\,,
\end{equation}
which leads to extra relations between $\RS$, $\TS$ and $\ES$ due to (\ref{offshellalgebra}). Moreover, it requires the introduction of several new tensors. In general, this process needs to be continued up to arbitrary order in the commutators, until it terminates. In this text, however, we will not go beyond first order since the most interesting properties of the algebra follow already from a single commutator on the fields.

\subsection{Zero modes \label{s:zeromodes}}
With the knowledge of the gauge generators from sections \ref{s:dewitt} and \ref{s:closure}, we can now address the (in)dependence of the gauge transformations. Again we will distinguish between two cases:
\begin{enumerate}
  \item For the formalism without an action, the question whether the gauge transformations $\{\delta(\Lambda),\,\delta(\Xi),\,\delta(\Phi)\}$ are (in)dependent, can be formulated as follows: do there exist vectors $\Zone{}^{a_0}{}_{a_1}$, such that for all $i$
\begin{equation}\label{zeromodes}
 \R^i{}_{a_0} \Zone{}^{a_0}{}_{a_1}=0\,?
\end{equation}
The index $a_1$ enumerates the possible outcomes. If (\ref{zeromodes}) has $m_1 \neq 0$ non-trivial solutions, then $a_1$ takes $m_1$ different values and it means that there exist $m_1$ dependencies between the gauge generators. In this case, the algebra is called reducible and the $\Zone{}^{a_0}{}_{a_1}$ are its zero modes. If (\ref{zeromodes}) has no non-trivial solutions, then the gauge transformations are independent and the algebra is called irreducible.
\item If the formalism is embedded into the framework with a classical action $S_0$, then we should consider the generators $\RS{}^i{}_{a_0}$ instead and equation (\ref{zeromodes}) has to be modified to
\begin{equation}\label{zeromodesS}
   \RS{}^i{}_{a_0} \ZSone{}^{a_0}{}_{a_1}=\partial_j S_0 \,\VSone{}^{ji}{}_{a_1}\,,
\end{equation}
for some tensors $\ZSone{}^{a_0}{}_{a_1}$ and $\VSone{}^{ij}{}_{a_1}=-\VSone{}^{ji}{}_{a_1}$.
The right hand side of (\ref{zeromodesS}) is now proportional to the field equations,  which means that the $\ZSone{}^{a_0}{}_{a_1}$ are \textit{on-shell} null vectors (or zero modes):
\begin{equation}
 \left.\RS{}^i{}_{a_0} \ZSone{}^{a_0}{}_{a_1}\right|_{\rm on-shell}=0\,.
\end{equation}
The presence of $\VSone{}^{ij}{}_{a_1}$ in (\ref{zeromodesS}) is a way to extend this statement off-shell. If $\{a_1\}$ is non-empty, then the gauge generators have $m_1$ on-shell dependencies and the algebra is reducible.
\end{enumerate}
The remainder of this section will be devoted to finding the solutions of (\ref{zeromodesS}). The solutions of (\ref{zeromodes}) are very similar and will be discussed in the summary at the end of this subsection.

The strategy to solve (\ref{zeromodesS}) will be to evaluate the different possibilities for the index $i$, and work out the summation over the $a_0$-index. Then we look for particular solutions which fix the precise form of the $a_1$-type indices. Let us first choose $i=\mu M$, then equation (\ref{zeromodesS}) becomes
\begin{equation}
\RS_{\mu}{}^M{}_Q \ZSone{}^Q{}_{a_1}+\RS_{\mu}{}^{M\,\nu}{}_{ \lc  RS \rf}\ZSone{}_{\,\nu}{}^{\lc  RS \rf}{}_{a_1}+\RS_{\mu}{}^{M\,\nu \rho}{}_{ \lc Q  \lc RS \rf\rf }\ZSone{}_{\,\nu \rho}{}^{ \lc Q \lc RS \rf\rf }{}_{a_1}=-\partial_j S_0 \,\VSone{}_{\,\mu}{}^{ M}{}^{j}{}_{a_1}\,.
\end{equation}
Using the fact that $\RS_{\mu}{}^{M\,\nu \rho}{}_{\lc  Q \lc RS\rf\rf  }=0$, $\RS_{\mu}{}^M{}_Q=\R_{\mu}{}^M{}_Q$ and $\RS_{\mu}{}^{M\,\nu}{}_{ \lc RS \rf}=\R_{\mu}{}^{M\,\nu}{}_{\lc  RS \rf}$, we get
\begin{equation}
D_{\mu}{}^{ M}{}_Q \ZSone{}^Q{}_{a_1} -Y^M{}_{RS} \ZSone{}_{\,\mu}{}^{ \lc RS \rf}{}_{a_1}=-\partial_j S_0\,\VSone{}_{\,\mu}{}^M{}^{j}{}_{a_1}\,. \label{muMzeromodes}
\end{equation}
A particular solution of this equation exists if we choose the $a_1$-index of the $\lc KL \rf$-type and
\begin{eqnarray}
\label{zeromodes1} \ZSone{}^{Q}{}_{\lc KL \rf}&=&Y^Q{}_{KL}\,, \hspace{1cm} \ZSone{}_{\mu}{}^{ \lc RS \rf }{}_{\lc KL \rf}\;=\;D_\mu{}^{RS}{}_{KL}\,,\\\label{solV1}
 \VSone{}_\mu{}^M{}^{j}{}_{\lc KL \rf}&=&0\quad \forall \,j\,.
\end{eqnarray}
Then (\ref{muMzeromodes}) reduces to
\begin{equation}
 D_{\mu}{}^{ M}{}_Q Y^Q{}_{NP} -Y^M{}_{RS} D_{\mu}{}^{ RS}{}_{NP}=0 \,,
\end{equation}
which is a property of the special derivatives (see equation (\ref{DY1})) and reflects the gauge invariance of the tensors $Y^Q{}_{NP}$.

In order to determine the remaining tensor $\ZSone{}_{\mu \nu}{}^{ \lc Q \lc RS \rf\rf }{}_{\lc KL\rf }$, we choose the $i$-index in (\ref{zeromodesS}) to be of the $\mu\nu \lc NP \rf$-type and $a_1=\lc KL\rf $:
\begin{eqnarray}
\nonumber &&\RS{}_{\mu\nu}{}^{\lc NP \rf}{}_Q \ZSone{}^Q{}_{\lc KL \rf}+\RS_{\mu\nu}{}^{\lc NP\rf }{}^{\rho}{}_{ \lc RS \rf}\ZSone{}_{\rho}{}^{\lc RS\rf }{}_{\lc KL \rf}+\RS_{\mu\nu}{}^{\lc NP\rf }{}^{\rho \sigma}{}_{\lc Q \lc RS \rf\rf }\ZSone{}_{\rho \sigma}{}^{ \lc Q \lc RS \rf\rf }{}_{\lc KL\rf }\\&&=-\VSone{}_{\mu\nu}{}^{\lc NP\rf\,j}{}_{ \lc KL\rf }\;\partial_j S_0\,.
\end{eqnarray}
Plugging in the generators (\ref{Bgenerators}), (\ref{generatorphi2}) and the zero modes (\ref{zeromodes1}) gives
\begin{eqnarray}
 \left(2A_{[\mu}{}^{\lc N}D_{\nu]}{}^{P \rf }{}_Q - 2\delta_Q{}^{\lc N}{\cal G}_{\mu\nu}{}^{ P\rf}\right)Y^Q{}_{KL} &&\\
 \nonumber+\left(2D_{[\mu}{}^{ NP}{}_{RS}-2A_{[\mu}{}^{ \lc N}Y^{P \rf}{}_{RS}\right)D_{\nu]}{}^{ RS}{}_{KL}&&\\
 \nonumber-Y^{NP}{}_{Q \lc RS \rf}\ZSone{}_{\mu\nu}{}^{\lc Q \lc RS\rf\rf}{}_{\lc KL\rf }&=&-\VSone{}_{\mu\nu}{}^{\lc NP\rf\,j}{}_{\lc KL \rf}\;\partial_j S_0\,.
\end{eqnarray}
Due to (\ref{DY1}), the first and fourth term on the left hand side cancel. To see how the remaining terms combine, we use the Ricci identity in the second line,\footnote{We have introduced the following shorthand notation here:
\begin{equation}
 X_{M \lc KL \rf}{}^{ NP }\equiv 2 X_{M \lc K}{}^{\lc N}\delta_{L\rf}{}^{P\rf}\,.
\end{equation} This definition can be generalized to arbitrary $X_{M_1 \lc M_2 \ldots M_p \rf .. \rf}{}^{\lc N_1\lc N_2 \ldots N_{p-1}\rf .. \rf}$. We refer to appendix \ref{a:covcontra} for more details.
}
\begin{equation}
 2D_{[\mu}{}^{ NP}{}_{RS}\,D_{\nu]}{}^{ RS}{}_{KL} = {\cal H}_{\mu\nu}{}^{M} X_{M \lc KL\rf }{}^{ NP } \,,
\end{equation}
and we get
\begin{eqnarray}
 \label{zeromintermediate1} - 2{\cal G}_{\mu\nu}{}^{\lc N}Y^{P \rf}{}_{KL}+{\cal H}_{\mu\nu}{}^{M} X_{M \lc KL \rf}{}^{ NP } \\
 \nonumber -\left(2\delta_Q{}^{\lc N}Y^{P\rf}{}_{RS} - X_{Q \lc RS \rf}{}^{ NP }\right)\ZSone{}{}_{\mu\nu}{}^{\lc Q \lc RS\rf\rf}{}_{\lc KL \rf}&=&-\VSone{}_{\mu\nu}{}^{ \lc NP\rf\,j}{}_{\lc KL \rf}\;\partial_j S_0\,.
\end{eqnarray}
The tensor structure in the first and second line look very similar and the most obvious choice is to combine the first and third term on the left hand side, as well as the second and fourth term. If we make the choice
\begin{equation}
 \ZSone{}{}_{\mu\nu}{}^{\lc Q \lc RS\rf\rf}{}_{ KL }= -{\cal G}_{\mu\nu}{}^{  \lc Q}{\mathbb P}^{RS\rf}{}_{KL}\,,
\end{equation}
equation (\ref{zeromintermediate1}) reduces to
\begin{equation}\label{V1}
 \left({\cal H}_{\mu\nu}{}^{M}-{\cal G}_{\mu\nu}{}^{M}\right) X_{M \lc KL \rf}{}^{ NP }
 =-\VSone{}_{\mu\nu}{}^{ NP\,j}{}_{ KL }\;\partial_j S_0\,.
\end{equation}
In order to determine the $\widetilde{\boldsymbol{V}}_{\!\!\!\boldsymbol{(1)}}$-tensors on the right hand side, we need to substitute the equations of motion $\partial_j S_0$, which have been given in (\ref{eomA}) and (\ref{eomB}):
\begin{eqnarray}
 \label{zeromintermediate2}\left({\cal H}-{\cal G}\right)_{\mu\nu}{}^{M} X_{M \lc KL \rf}{}^{ NP } &=&-\VSone{}_{\mu\nu}{}^{\lc NP\rf}{}_{\rho}{}^{R}{}_{\lc KL \rf}\left(\frac{\partial S_0}{\partial A_\rho{}^R}\right)
 \\\nonumber&&-\VSone{}_{\mu\nu}{}^{\lc NP\rf}{}_{\rho\sigma}{}^{\lc RS\rf}{}_{\lc KL\rf }\left(\frac{1}{4} \varepsilon^{\rho\sigma\lambda\tau}\Omega_{MQ}Y^M{}_{RS}\left({\cal H}-{\cal G}\right)_{\lambda\tau}{}^Q\right)\,.
\end{eqnarray}
The first term on the right hand side vanishes since also $\VSone{}_{\rho}{}^{R}{}_{\mu\nu}{}^{ \lc NP\rf}{}_{\lc KL \rf}=0$ (see (\ref{solV1})) and because $\VSone{}^{ij}{}_{\lc KL\rf}$ is antisymmetric in $[ij]$. This leaves us with two terms that are both proportional to $({\cal H}-{\cal G})$. It requires some calculational effort (and the linear constraint (\ref{linearconstr}) on the embedding tensor) to show that equation (\ref{zeromintermediate2}) is satisfied if we choose
\begin{equation}
 \VSone{}_{\mu\nu}{}^{\lc NP\rf}{}_{\rho\sigma}{}^{\lc RS\rf}{}_{\lc KL\rf }=-4\varepsilon_{\mu\nu\rho\sigma}\delta_{ \lc K}{}^{ \lc N}\Omega^{P\rf\lc R}\delta_{L\rf }{}^{S\rf }\,.
\end{equation}
To summarize, we have shown that
\begin{equation}
 \ZSone{}^{a_0}{}_{\lc KL \rf}=\left(\begin{array}{c}
                                    Y^Q{}_{KL}\\D_\mu{}^{RS}{}_{KL}\\
-{\cal G}_{\mu\nu}{}^{ \lc Q}{\mathbb P}^{RS\rf}{}_{KL}
                                   \end{array}
\right) \quad\mbox{and} \quad \VSone{}^{ij}{}_{\lc KL\rf}=\left(\begin{array}{cc}
                                    0&0\\0&-4\varepsilon_{\mu\nu\rho\sigma}\delta_{\lc K}{}^{ \lc N}\Omega^{P\rf\lc R}\delta_{L\rf }{}^{S\rf }
                                   \end{array}\right)\,
\end{equation}
form a non-trivial solution of (\ref{zeromodesS}) for $a_1= \lc KL \rf$.

However, this is not the only value of $a_1$ for which a solution can be found. A calculation which is very similar to the one above, reveals that there are two more solutions, namely for ${a_1}=\rho \lc K \lc LM \rf\rf$ and ${a_1}=\rho\sigma \lc K \lc L \lc MN \rf\rf\rf$. In both cases, all the $\VSone{}^{ij}{}_{a_1}$ are zero, and
\begin{equation}
 \ZSone{}^{a_0\,\rho}{}_{\lc K \lc LM \rf\rf}=\left(\begin{array}{c}
0\\
 Y^{RS}{}_{ K \lc LM \rf}\delta_\mu^\rho\\
 2D_{[\mu}{}^{Q\lc RS\rf}{}_{K\lc LM\rf}\delta_{\nu]}^\rho
\end{array}\right)\,,\quad \ZSone{}^{a_0\,\rho\sigma}{}_{\lc K \lc L \lc MN \rf\rf\rf}=\left(\begin{array}{c}
0\\
0\\
 Y^{Q \lc RS\rf}{}_{K \lc L \lc MN \rf\rf}\delta^{[\rho\sigma]}_{[\mu\nu]}
\end{array}\right).
\end{equation}
In the expression for the third zero mode appears a tensor $Y^{Q \lc RS\rf}{}_{K \lc L \lc MN \rf\rf}$, which is a generalization of $Y^{Q}{}_{K  L}$ and $Y^{RS}{}_{K \lc LM\rf}$. It is defined\footnote{In general, we will introduce tensors $Y^{M_1\lc \hdots M_{p-1}\rf..\rf}{}_{K_1\lc \hdots K_p\rf..\rf}$ that vanish upon contraction with $Y^{K_1\lc \hdots K_p\rf..\rf}{}_{N_1\lc \hdots N_{p+1}\rf..\rf}$, and their precise definition is given in (\ref{Yp}).} in (\ref{defY3}) and has the property that
\begin{equation}
Y^{TU}{}_{Q\lc RS\rf} Y^{Q \lc RS\rf}{}_{K \lc L \lc MN\rf\rf}=0\,.
\end{equation}

\paragraph{Summary and discussion:} The generators that make up the gauge algebra in the $D=4$ embedding tensor formalism with action $S_0$, are not all independent. For the indices $i$ and $a_0$ restricted to $\{\mu N,\,\mu\nu \lc NP\rf\}$ and $\{Q,\,\mu \lc RS \rf,\,\mu\nu \lc Q \lc RS\rf\rf\}$ respectively, we checked that equation (\ref{zeromodesS}) has non-trivial solutions for the zero modes $\ZSone{}^{a_0}{}_{a_1}$ and corresponding tensors $\VSone{}^{ij}{}_{a_1}$. We found $3$ solutions in total, more precisely for $a_1= \lc KL\rf$, $a_1=\rho \lc K \lc LM\rf\rf$ and $a_1=\rho\sigma \lc K \lc L \lc MN\rf\rf\rf$. Then $\ZSone{}^{a_0}{}_{a_1}$ is a $(3 \times 3)$ block matrix; the rows are enumerated by $a_0$ and the columns by $a_1$:
\begin{equation}\label{solzeromodes}
\ZSone{}^{a_0}{}_{a_1}= \left(
\begin{array}{ccc}
Y^Q{}_{KL}&0&0\\
D_\mu{}^{RS}{}_{KL}&Y^{RS}{}_{K \lc LM \rf}\delta_\mu^\rho&0\\
-{\cal G}_{\mu\nu}{}^{ \lc Q}{\mathbb P}^{RS\rf}{}_{KL}&2D_{[\mu}{}^{Q\lc RS\rf}{}_{K\lc LM\rf}\delta_{\nu]}^\rho&Y^{Q \lc RS\rf}{}_{K \lc L \lc MN\rf\rf }\delta^{\rho\sigma}_{[\mu\nu]}
\end{array}
\right).
\end{equation}
We recognize a certain systematics in this solution: the diagonal entries are all proportional to a $Y$-tensor and the 21- and 32-elements contain a derivative. This special structure will be further investigated in the next section, where we show that $\ZSone{}^{a_0}{}_{a_1}$ has non-maximal rank, which means that not all zero modes are independent.

Finally, we remark here that the solutions $\Zone{}^{a_0}{}_{a_1}$ of (\ref{zeromodes}) are identical, except for the lower left entry, where ${\cal G}_{\mu\nu}{}^Q$ should be replaced by ${\cal H}_{\mu\nu}{}^Q$.

\subsection{Higher stage zero modes \label{s:higherstagezerom}}
If the $3$ solutions for $\Zone{}^{a_0}{}_{a_1}$ or $\ZSone{}^{a_0}{}_{a_1}$ are independent, then the theory is called first-stage reducible. However, this may not happen; there can be `level-two' gauge invariances that reflect the dependencies among the $\Zone{}^{a_0}{}_{a_1}$ or $\ZSone{}^{a_0}{}_{a_1}$. This reasoning can be repeated for the level-two generators, and it possibly leads to dependencies at higher stages. This brings us to the concept of an $L$-th stage reducible theory, which means that only at level $L$, all the generators are independent. In order to determine the level $L$ for the gauge structure of the embedding tensor formalism, we will investigate the dependencies among the zero modes in (\ref{solzeromodes}). We need to solve an equation that is similar to (\ref{zeromodes}) or (\ref{zeromodesS}):
 \begin{enumerate}
   \item  For the transformations in the absence of an action, we look for non-trivial tensors $\Ztwo{}^{a_1}{}_{a_2}$ that are solutions of
\begin{equation}\label{zeromodes2}
 \Zone{}^{a_0}{}_{a_1} \Ztwo{}^{a_1}{}_{a_2} =0\,.
\end{equation}
The index $a_2$ labels the $m_2$ different solutions and therefore the possible dependencies of the zero modes. The new tensors $\Ztwo{}^{a_1}{}_{a_2}$ are called `zero modes for zero modes' or second stage zero modes.
\item In the presence of an action, equation (\ref{zeromodes2}) needs to be modified to
\begin{equation}\label{zeromodesS2}
\ZSone{}^{a_0}{}_{a_1} \ZStwo{}^{a_1}{}_{a_2} = \partial_i S_0\; \VStwo{}^{i\, a_0}{}_{a_2}\,.
 \end{equation}
The $\ZStwo{}^{a_1}{}_{a_2}$ are $m_2$ on-shell null vectors of the zero modes. The tensors $\VStwo{}^{i\, a_0}{}_{a_2}$ in (\ref{zeromodesS2}) provide an off-shell extension of this statement.
 \end{enumerate}
We will look for non-trivial solutions of (\ref{zeromodesS2}) with $\ZSone{}^{a_0}{}_{a_1}$ given in (\ref{solzeromodes}).\footnote{The solutions of (\ref{zeromodes2}) will again be very similar, and can be found at the end of this section.} Our strategy will be to make a motivated guess for the solutions, and then check that (\ref{zeromodesS2}) is indeed satisfied. From the previous section, we know that $a_0 \in \{K_1,\, \mu \lc K_1K_2\rf,\, \mu\nu \lc K_1\lc K_2K_3\rf\rf\}$ and $a_1 \in \{\lc K_1K_2\rf,\, \mu \lc K_1\lc K_2K_3\rf\rf,\, \mu\nu \lc K_1\hdots K_4\rf..\rf\}$. Comparing these two index sets, we expect that this structure can be continued and
\begin{equation}
 a_2 \in \{\lc K_1\lc K_2 K_3\rf\rf,\, \mu \lc K_1\hdots K_4\rf..\rf,\, \mu\nu \lc K_1\hdots K_5\rf..\rf\}\,.
\end{equation}
Therefore, we propose the following form for $\ZStwo{}^{a_1}{}_{a_2}$, which looks very similar to the expression for the zero modes in (\ref{solzeromodes}):
\begin{equation}\label{solzeromodesS2}
  \ZStwo{}^{a_1}{}_{a_2}=\left( \begin{array}{ccc}
   -Y^{M_1M_2}{}_{K_1\lc K_2K_3\rf}&0&0\\
   D_{\mu}{}^{ M_1\lc M_2M_3\rf}{}_{K_1\lc K_2K_3\rf}& -Y^{M_1 \lc M_2M_3\rf }{}_{K_1 \lc \hdots K_4\rf\rf}\delta_\mu^{\rho}&0\\
  {\cal G}_{\mu\nu}{}^{\lc M_1}{\mathbb P}^{\hdots M_4\rf\rf}{}_{K_1\lc K_2K_3\rf} & 2 D_{[\mu}{}^{ M_1
   \lc \hdots M_4\rf\rf}{}_{K_1 \lc \hdots K_4\rf\rf}\delta_{\nu]}^\rho&-Y^{M_1 \lc \hdots M_4\rf\rf}{}_{K_1 \lc \hdots K_5\rf..\rf}\delta_{[\mu\nu]}^{[\rho\sigma]}
  \end{array} \right).
 \end{equation}
On the diagonal, $Y$-tensors appear with an extra minus sign compared to the expression for the zero modes (\ref{solzeromodes}). The 21- and 32-elements contain a derivative and the lower left entry is proportional to the scalar dependent field strength.

Let us now compute the different entries in the matrix product $\ZSone{}^{a_0}{}_{a_1} \ZStwo{}^{a_1}{}_{a_2}$ and show that they are proportional to the field equations, just as in (\ref{zeromodesS2}). This calculation is a check on the validity of (\ref{solzeromodesS2}) and it gives the correct expression for $\VStwo{}^{i\, a_0}{}_{a_2}$.

We start with the computation of the $11$-element, which corresponds to multiplying the first row of $\ZSone$ with the first column of $\ZStwo$:
\begin{equation}
\left[ \ZSone{}^{a_0}{}_{a_1} \ZStwo{}^{a_1}{}_{a_2}\right]_{11}= - Y^{M_1}{}_{N_1N_2}Y^{N_1N_2}{}_{K_1 \lc K_2K_3\rf}\,.
\end{equation}
This expression vanishes because of the orthogonality of the $Y$-tensors, see (\ref{YY1}). It also means that (\ref{zeromodesS2}) is satisfied iff $\VStwo{}^{i\, M_1}{}_{\lc K_1 \lc K_2K_3\rf\rf}=0$ for all values of $i$.
A similar reasoning can be made for the other diagonal elements in the matrix product: we use (\ref{YYp}) to show that they vanish and this is consistent with (\ref{zeromodesS2}) iff the corresponding $\VStwo{}^{i\, a_0}{}_{a_2}$'s also vanish.
The $12$-, $13$- and $23$-entries are trivially satisfied, which brings us to the $21$ and $32$ elements:
\begin{eqnarray}
\nonumber\left[ \ZSone{}^{a_0}{}_{a_1} \ZStwo{}^{a_1}{}_{a_2}\right]_{21}&=&-D_{\mu}{}^{M_1M_2}{}_{N_1N_2}Y^{N_1N_2}{}_{K_1\lc K_2K_3\rf}+Y^{M_1M_2}{}_{N_1\lc N_2N_3\rf}D_\mu{}^{N_1\lc N_2N_3\rf}{}_{K_1\lc K_2K_3\rf}\,,\\
\nonumber \left[ \ZSone{}^{a_0}{}_{a_1} \ZStwo{}^{a_1}{}_{a_2}\right]_{32}&=&-2D_{[\mu}{}^{M_1\lc M_2 M_3\rf}{}_{N_1\lc N_2 N_3\rf}\delta_{\nu]}^\rho Y^{N_1\lc N_2 N_3\rf}{}_{K_1\lc\hdots K_4\rf\rf}\\&&+2Y^{M_1\lc M_2 M_3\rf}{}_{N_1\lc\hdots N_4\rf\rf}D_{[\mu}{}^{N_1\lc \hdots N_4\rf\rf}{}_{K_1\lc \hdots K_4\rf\rf}\delta_{\nu]}^{\rho} \,.
\end{eqnarray}
Both expressions vanish due to the properties of the covariant derivative, see (\ref{DYp}). This means that, in order to satisfy (\ref{zeromodesS2}), also $\VStwo{}^{i}{}_{\mu}{}^{\lc M_1M_2\rf}{}_{\lc K_1\hdots K_3\rf\rf}=\VStwo{}^{i}{}_{\mu\nu}{}^{\lc M_1\hdots M_3\rf\rf\,\rho}{}_{\lc K_1\hdots K_4\rf..\rf}=0$ for all values of $i$. Finally, we consider the $31$ element:
\begin{eqnarray}
 \nonumber \left[\ZSone{}^{a_0}{}_{a_1} \ZStwo{}^{a_1}{}_{a_2}\right]_{31} &=&  {\cal G}_{\mu\nu}{}^{ \lc {M_1}}{\mathbb P}^{{M_2}{M_3} \rf}{}_{N_1N_2} Y^{N_1N_2}{}_{K_1\lc K_2K_3\rf}\\
 \nonumber &&+2 D_{[\mu}{}^{ {M_1}\lc {M_2}{M_3}\rf}{}_{N_1 \lc N_2N_3\rf} D_{\nu]}{}^{ N_1 \lc N_2N_3\rf}{}_{K_1\lc K_2K_3\rf}\\
 \nonumber&&+ Y^{{M_1} \lc {M_2}{M_3} \rf}{}_{N_1 \lc \hdots N_4\rf\rf}{\cal G}_{\mu\nu}{}^{ \lc N_1}{\mathbb P}^{\hdots N_4\rf\rf}{}_{K_1\lc K_2K_3\rf}\\
 \nonumber&=& {\cal G}_{\mu\nu}{}^{ N_1} \delta_{N_1}{}^{\lc {M_1}}Y^{{M_2}{M_3}\rf}{}_{K_1\lc K_2K_3\rf}+{\cal H}_{\mu\nu}{}^{ N_1} X_{N_1 \lc K_1\lc K_2K_3\rf\rf}{}^{\lc M_1 \lc {M_2}{M_3} \rf \rf}\\
 \nonumber&& + {\cal G}_{\mu\nu}{}^{ N_1}\left(-\delta_{N_1}{}^{\lc {M_1}} Y^{{M_2}{M_3} \rf}{}_{K_1\lc K_2K_3\rf}-X_{N_1 \lc K_1\lc K_2K_3\rf\rf}{}^{\lc {M_1}\lc {M_2}{M_3} \rf \rf}\right)\\
 \label{stage2zeromod}&=&\left({\cal H}-{\cal G}\right)_{\mu\nu}{}^{ N_1} X_{N_1 \lc K_1\lc K_2K_3\rf\rf}{}^{\lc {M_1}\lc {M_2}{M_3} \rf \rf}\,.
 \end{eqnarray}
For the second equality, we used the Ricci identity
\begin{equation}
 2 D_{[\mu}{}^{ {M_1}\lc {M_2}{M_3}\rf}{}_{N_1 \lc N_2N_3\rf}D_{\nu]}{}^{N_1 \lc N_2N_3\rf}{}_{K_1\lc K_2K_3\rf}={\cal H}_{\mu\nu}{}^{N_1} X_{N_1 \lc K_1\lc K_2K_3\rf\rf}{}^{\lc {M_1} \lc {M_2}{M_3} \rf \rf}\,,
\end{equation}
and the definition (\ref{Y2}).
The result (\ref{stage2zeromod}) is the analog of (\ref{V1}), and in order to satisfy (\ref{zeromodesS2}), we have
\begin{equation}
 \left({\cal H}-{\cal G}\right)_{\mu\nu}{}^{ N_1} X_{N_1 \lc K_1\lc K_2K_3\rf\rf}{}^{\lc {M_1}\lc {M_2}{M_3} \rf \rf}=\frac{\partial S_0}{\partial B_{\rho\sigma}{}^{\lc  N_1N_2\rf}}\VStwo{}_{\rho\sigma}{}^{\lc N_1N_2 \rf}{}_{\mu\nu}{}^{\lc {M_1}\lc {M_2}{M_3} \rf\rf}{}_{\lc K_1\lc K_2K_3\rf\rf}\,.
\end{equation}
This defines the tensor $\VStwo$ which can be determined via a short calculation that requires the use of the linear constraint (\ref{linearconstr}) on the embedding tensor. We find that
\begin{eqnarray}
  \VStwo{}_{\rho\sigma}{}^{\lc N_1N_2 \rf}{}_{\mu\nu}{}^{\lc {M_1}\lc {M_2}{M_3} \rf\rf}{}_{\lc K_1\lc K_2K_3 \rf\rf}&=&-2\varepsilon_{\mu\nu\rho\sigma}\Big{(}\delta_{\lc K_1}{}^{\lc N_1}\Omega^{N_2\rf P_1}\delta_{\lc K_2}{}^{P_2}\delta_{K_3\rf\rf}{}^{P_3}\\
\nonumber&&\left.\qquad\qquad+\delta_{\lc K_1}{}^{P_1}\delta_{\lc K_2}{}^{\lc N_1}\Omega^{N_2\rf P_2}\delta_{K_3\rf\rf}{}^{P_3}\right.\\\nonumber
 &&\qquad\qquad+\delta_{\lc K_1}{}^{P_1}\delta_{\lc K_2}{}^{P_2}\delta_{K_3 \rf\rf}{}^{\lc N_1}\Omega^{N_2\rf P_3}
 \Big{)}{\mathbb P}^{{M_1}\lc {M_2}{M_3} \rf}{}_{P_1\lc P_2P_3\rf}\,.
\end{eqnarray}
So in the end, we have proven that our proposal for $\ZStwo{}^{a_1}{}_{a_2}$ in (\ref{solzeromodesS2}) is indeed a solution of (\ref{zeromodesS2}) for each value of $a_2$. It means that the zero modes $\ZSone{}^{a_0}{}_{a_1}$ are not all independent and the gauge algebra is at least reducible up to level $2$. The same conclusion holds for the zero modes in the embedding tensor formalism without an action, i.e. $\Zone{}^{a_0}{}_{a_1}$. These are also not independent and the solutions of (\ref{zeromodes2}) are identical to (\ref{solzeromodesS2}), except for the lower left corner, where the field strengths ${\cal G}_{\mu\nu}{}^M$ should be replaced by ${\cal H}_{\mu\nu}{}^M$.

Given these non-trivial expressions for the 1st and 2nd stage zero modes, one could wonder whether there exists a level for which this construction terminates. In other words, is there a level $s$, for which
\begin{equation}\label{arbitrarystagezeromod}
 {\widetilde{\boldsymbol{Z}}}_{\!\!\!\boldsymbol{(s-1)}}{}^{a_{s-2}}{}_{a_{s-1}}{\widetilde{\boldsymbol{Z}}}_{\!\!\!\boldsymbol{(s)}}{}^{a_{s-1}}{}_{a_s}=\partial_i S_0\,{\widetilde{\boldsymbol{V}}}_{\!\!\!\boldsymbol{(s)}}{}^{i\, a_{s-2}}{}_{a_s}
\end{equation}
has \textit{no} non-trivial solutions? If this is the case for a value $s=L$, then the theory is called $L$-th stage reducible.

What about the embedding tensor formalism? Is it finitely reducible? A priori, there does not seem to be a level at which the above construction comes to an end. Indeed, one can propose the following expressions for the zero modes and non-vanishing ${\widetilde{\boldsymbol{V}}}$-tensors at arbitrary level $s \geq 1$:
\begin{equation}\label{Zgeneralstage}
 {\widetilde{\boldsymbol{Z}}}_{\!\!\!\boldsymbol{(s)}}{}^{a_{s-1}}{}_{a_s}=\left[\begin{array}{ccc}
  [A_{(s)}]&[B_{(s)}]&[C_{(s)}]
 \end{array}\right]\,,
\end{equation}
with
\begin{equation}
 [A_{(s)}]=\left[\begin{array}{c}
  (-1)^{s+1}\,Y^{M_1 \lc M_2 \hdots M_s\rf..\rf}{}_{N_1 \lc N_2 \hdots N_{s+1}\rf..\rf}\\
  D_{\mu}{}^{ \,M_1 \lc M_2\hdots M_{s+1}\rf..\rf}{}_{N_1 \lc N_2 \hdots N_{s+1}\rf..\rf}\\
  (-1)^{s}\,{\cal G}_{\mu\nu}{}^{ \lc M_1}{\mathbb P}^{M_2 \lc\hdots  M_{s+2}\rf..\rf}{}_{N_1 \lc N_2 \hdots N_{s+1}\rf..\rf}
 \end{array}\right]\,,
\end{equation}
\begin{equation}
 [B_{(s)}]=\left[\begin{array}{c}
 0\\
  (-1)^{s+1}\,Y^{M_1 \lc M_2 \hdots M_{s+1}\rf..\rf}{}_{N_1 \lc N_2 \hdots N_{s+2}\rf.. \rf}\delta^\rho_\mu\\
  2 D_{[\mu}{}^{ M_1\lc M_2 \hdots  M_{s+2} \rf..\rf}{}_{N_1\lc N_2 \hdots N_{s+2}\rf..\rf}\delta_{\nu]}^\rho
 \end{array}\right]\,,
\end{equation}
\begin{equation}
 [C_{(s)}]=\left[\begin{array}{c}
  0\\
  0\\
  (-1)^{s+1}\,Y^{M_1 \lc M_2 \hdots M_{s+2}\rf.. \rf}{}_{N_1 \lc N_2 \hdots N_{s+3}\rf.. \rf}\delta^{[\rho\sigma]}_{[\mu\nu]}
 \end{array}\right]\,,
\end{equation}
and
\begin{eqnarray}\label{Vgeneralstage}
 {\widetilde{\boldsymbol{V}}}_{\!\!\!\boldsymbol{(s)}}{}_{\rho\sigma}{}^{\lc N_1N_2\rf}{}_{\mu\nu}{}^{M_0\lc M_1\hdots M_s\rf..\rf}{}_{K_0\lc K_1\hdots K_s\rf..\rf}&=&-2\varepsilon_{\mu\nu\rho\sigma}\Big{(}\delta_{\lc K_0}{}^{\lc N_1}\Omega^{N_2\rf P_0}\delta_{\lc K_1}{}^{P_1}\hdots \delta_{K_s \rf..\rf}{}^{P_s}\\
 \nonumber&&+\,\delta_{\lc K_0}{}^{P_0}\delta_{\lc K_1}{}^{\lc N_1}\Omega^{N_2\rf P_1}\delta_{\lc K_2}{}^{P_2}\hdots \delta_{K_s \rf..\rf}{}^{P_s}\\
 \nonumber&&+\,\hdots\\
 \nonumber&&\,+\,\delta_{\lc K_0}{}^{P_0}\hdots\delta_{K_s \rf..\rf}{}^{\lc N_1}\Omega^{N_2\rf P_s} \Big{)}{\mathbb P}^{M_0\lc M_1 \hdots M_s\rf..\rf}{}_{P_0\lc P_1 \hdots P_s\rf..\rf}\,.
\end{eqnarray}
Equation (\ref{arbitrarystagezeromod}) is always satisfied for this combination of tensors, irrespective of the value of $s$. Therefore, we conclude that there always exists a zero mode at every arbitrary level and the theory is infinitely reducible.

Of course, this is just a formal statement since in particular examples, one needs to evaluate the different projection operators for the special brackets in (\ref{Zgeneralstage})-(\ref{Vgeneralstage}). For certain choices of the embedding tensor, the projectors ${\mathbb P}^{M_1\lc M_2 \hdots M_p\rf..\rf}{}_{N_1\lc N_2 \hdots N_p\rf..\rf}$ might vanish for $p$ bigger than a certain value, say $\ell$. This means that also the corresponding objects in (\ref{Zgeneralstage})-(\ref{Vgeneralstage}) with more than $\ell$ upper or lower indices are identically zero. Therefore, $\ell$ determines the level, $L$, at which the zero modes of the algebra become independent. We conclude that a case-by-case study is needed to determine $L$ and as such, no general statement can be made about its value.

\vspace{2mm}
We have now come to the end of our discussion on the gauge structure of the embedding tensor formalism with 1- and 2-forms and local transformations $\delta(\Lambda)$, $\delta(\Xi)$ and $\delta(\Phi)$. We found an algebra that is open in general, with field dependent structure functions and a hierarchy of zero modes that has no obvious ending. The details of this gauge structure are contained in a large set of tensors, such as the gauge generators, structure functions, zero modes, etc. These are complicated expressions of the fields and the embedding tensor, which makes it hard to take them into account in explicit calculations.  Therefore, one might wonder whether there exists an underlying prescription that provides a unified picture for these complicated tensors. In the next section we will see that such a unifying formalism does exist and that all the gauge structure tensors naturally fit into one `master equation'.

\section{BV formalism \label{s:BV}}
The formalism that we have in mind is the field-antifield or Batalin-Vilkovisky (BV) formalism. From the introduction we recall that this formalism was originally introduced as an extension of the Faddeev-Popov procedure to quantize a broader class of field theories with local symmetries. It is particularly useful for theories with a complicated gauge structure such as open, soft and/or reducible algebras. In the previous sections we saw that the embedding tensor formalism falls into this class and the BV formalism therefore provides all the tools for its quantization. However, we will not pursue this quantization, but rather concentrate on how the (classical) embedding tensor formalism fits into the structure of the classical BV formalism.

To this end, we introduce in section \ref{s:classicalBV} all the ingredients that make up the classical BV formalism. Then, in section \ref{s:embeddingBV}, we will see how the embedding tensor formalism fits into this framework and how the BV formalism provides a simplified description for its complicated tensor structure.

\subsection{Classical BV theory \label{s:classicalBV}}
Consider a classical system described by the action\footnote{In the previous sections about the embedding tensor formalism, we distinguished between the gauge algebra in the presence and in the absence of an action. To introduce the BV formalism we start with a system determined by an action $S_0$. Once we have introduced the formalism it is easy to consider the case where there is no gauge-invariant classical action.}
$S_0[\phi]$ that is a functional of the bosonic fields $\phi^i$. This means that the fields have even parity, i.e.
\begin{equation}
\epsilon[\phi^i]=0\,.
\end{equation}
In general the classical action $S_0$ can also contain fermionic degrees of freedom, but this case will not be considered here.

The theory has $m_0$ bosonic gauge symmetries that are generated by $\RS{}^i{}_{a_0}$ and have corresponding local parameters $\varepsilon^{a_0}$. This then leads to $m_0$ Noether identities as in (\ref{Noether}), an expression for the gauge commutators as in (\ref{offshellalgebra}), $m_1$ zero modes as solutions of (\ref{zeromodesS}), etc.
All these equations are written down in terms of certain tensors that determine the complete gauge structure of the theory. The main purpose of the classical BV-formalism is to provide a consistent framework that incorporates all these tensors in a transparent way. In particular, this is achieved through the construction of a new action, denoted by $S$, which is an extension of the classical action $S_0$. In brief, the construction of $S$ involves five steps, each of which will be discussed in more detail later on.
\begin{enumerate}
	\item Ghost fields are introduced to compensate for the gauge degrees of freedom. When dealing with a reducible system (in which the gauge transformations are not all independent),
	      also higher stage ghost fields need to be introduced. The original configuration space, consisting of the $\phi^i$, is enlarged to include these ghost fields, ghosts for ghosts, etc..
	\item For each field, thus also for the (higher stage) ghost fields, an antifield is introduced.
	\item On the space of fields and antifields, one defines an odd symplectic structure $(\,.\,,\,.\,)$, called the antibracket.
	\item The classical action $S_0$ is extended to include terms involving fields and antifields and is denoted by $S$. It has to satisfy certain boundary conditions, such as the requirement that in the limit where all antifields are put to zero, the extended action $S$ reduces to $S_0$.
\item Finally, one imposes the classical master equation, $(S,S)=0$.	One finds solutions $S$ to this equation, subject to the appropriate boundary conditions. It turns out that these solutions are an expansion in the antifields and that the coefficients in the expansion are exactly the tensors that determine the gauge structure of the theory.
\end{enumerate}
Let us now consider each of these steps in more detail.

\paragraph{Ghosts.}
Suppose we are dealing with an irreducible theory with $m_0$ gauge invariances and corresponding parameters $\varepsilon^{a_0}$. Then at the quantum level $m_0$ ghost fields $c_{(0)}{}^{a_0}$ are needed, i.e. one for each parameter. However, for our purposes it is useful to introduce these ghost fields already at the classical level. Hence, the complete set of classical fields is $\chi^n=\{\phi^i, c_{(0)}{}^{a_0} \}$.

In a reducible theory the $m_0$ gauge invariances are not all independent; there exist zero modes for the gauge invariances. In principle these zero modes imply that
we have introduced too many gauge parameters, but that can be necessary in order to preserve the covariance or locality of the theory. If there are $m_1$ first-level
zero modes then one adds the ghost-for-ghost fields $c_{(1)}{}^{a_1}$ ($a_1=1, \cdots, m_1$) to the above set $\chi^n$. In general for an $L$-stage reducible theory, the set of fields
$\chi^n$ (where $n=1, \cdots, N$) is
\begin{equation}
\chi^n=\left\{\phi^i, c_{(s)}{}^{a_s};\, s=0, \cdots, L\,;\, a_s=1, \cdots, m_s\right\}.
\end{equation}
The ghosts are defined as having opposite statistics to the corresponding gauge parameter, ghost for ghosts as having the same statistics as the gauge parameter, and
so on, with the statistics alternating for higher level ghosts. We can write this as
\begin{equation}
\epsilon[c_{(s)}{}^{a_s}]=(s+1)\mbox{mod } 2,
\end{equation}
where $\epsilon[c_{(s)}{}^{a_s}]$ denotes the parity of the (higher stage) ghost.
Moreover, an additive conserved charge, called ghost number $\mathop{\rm{gh}}[\chi^n]$, is assigned to each of these fields $\chi^n$. The classical fields $\phi^i$ have ghost number zero, whereas ordinary ghosts have ghost number one. Ghost for ghosts (first level ghosts), have ghost number two etc.
\begin{equation}
\mbox{gh}[\phi^i]=0\,,\qquad\mathop{\rm{gh}}[c_{(s)}{}^{a_s}]=s+1\,.
\end{equation}

\paragraph{Antifields.}
Next, one introduces an antifield $\chi^*_n$ ($n=1, \cdots, N$) for each field $\chi^n$. These antifields should be thought of as a mathematical tool to set up the formalism. The ghost number and statistics of $\chi^*_n$ are
\begin{eqnarray}
\label{ghantifield}\mathop{\rm{gh}}[\chi^*_n]&=&-\mathop{\rm{gh}}[\chi^n]-1\,, \\
\label{parantifield}\epsilon[\chi^*_n]&=&\left(\epsilon[\chi^n]+1\right)\mbox{mod }2\,,
\end{eqnarray}
such that $\chi^n$ and $\chi^*_n$ have opposite statistics. In the future, we will denote the total set of fields and antifields\footnote{In order to refer to the fields and antifields simultaneously, we will use the terminology `Fields', with a capital letter F.} with $z^a=\{\chi^n, \chi^*_n\}$.
For each Field, we introduce an antifield number $afn[z^a]$ which will become important later on.
\begin{equation}\label{afn}
afn[z^a]=\begin{cases}
0:& \mbox{gh}[z^a] \ge 0\,, \\
-\mbox{gh}[z^a]:& \mbox{gh}[z^a] < 0\,.
\end{cases}
\end{equation}

\paragraph{The antibracket.}
On the space of Fields one introduces an odd symplectic structure, the antibracket $(\,.\,,\,.\,)$. It is defined by
\begin{equation}\label{antibracket}
(X,Y) \equiv \frac{\partial_r X}{\partial \chi^n}\frac{\partial_l Y}{\partial \chi^*_n}-\frac{\partial_r X}{\partial \chi^*_n}\frac{\partial_l Y}{\partial \chi^n},
\end{equation}
where the subscripts $\partial_r$ and $\partial_l$ denote right and left differentiation respectively, and $X$ and $Y$ are arbitrary functionals of the Fields $z^a$. In the case where $X$ and $Y$ are bosonic quantities, the antibracket has the following useful properties. It is similar to the Poisson bracket, but symmetric under the exchange of $X$ and $Y$. It has odd statistics, i.e. $\epsilon[(X,Y)]=1$. Moreover, the bracket of two identical bosonic functionals $X$ of the Fields is
\begin{equation}
(X,X)=2\frac{\partial_r X}{\partial \chi^n}\frac{\partial_l X}{\partial \chi^*_n}. \label{bosonfunctional}
\end{equation}
Finally we note that the definition in (\ref{antibracket}) can also be written as
\begin{equation}
 (X,Y)=\frac{\partial_r X}{\partial z^a} \Omega^{ab} \frac{\partial_l Y}{\partial z^b}\,, \qquad \mbox{where}\quad \Omega^{ab}\equiv
\left(\begin{matrix}
0&\delta_m^n\\
-\delta_m^n&0
\end{matrix}\right),
\end{equation}
which is why we call the antibracket a symplectic structure.

\paragraph{The extended action and boundary conditions.}
Let $S[\chi,\chi^*]$ be an arbitrary functional of the Fields with the dimension of an action, even parity $\epsilon[S]=0$, and zero ghost number $\mathop{\rm{gh}}[S]=0$. This functional is called an \textit{extended action} if it satisfies the following boundary conditions:
\begin{itemize}
 \item[(i)] In the `classical limit', $S$ reduces to $S_0$,
\begin{equation}
\left.S[\chi,\chi^*]\right|_{\chi^*_n=0}\;=\;S_0[\phi^i]\,, \label{classicallimit}
\end{equation}
i.e. when all the antifields are put to zero, the extended action reduces to the original action $S_0$.
This requirement means that $S$ can be written as an expansion in the antifields, with the classical action $S_0$ at zeroth order:
\begin{equation}
 S[\chi,\chi^*]=S_0 + \mbox{terms that are linear, quadratic,... in the antifields.}
\end{equation}
This can be made more precise if we order all the terms in $S$ according to their antifield number:
\begin{equation}\label{expansion}
S=\sum_kS_k=S_0+S_1+S_2+\cdots,
\end{equation}
where $afn(S_i)=i$. An expression for $S_i$ for the lowest orders of $i$ will be given in due course.

\item[(ii)] The second boundary condition that should be satisfied is more technical. It is called the properness condition and it takes the following form:
\begin{equation}\label{properness}
\left.\mbox{rank}\frac{\partial_l \partial_r S}{\partial z^a \partial z^b}\right|_{\Sigma}=N,
\end{equation}
where $N$ is the number of fields $\chi^n$ or antifields $\chi^*_n$ and $\Sigma$ denotes the subspace of stationary points in the space of Fields,
\begin{equation}
z^a_0 \in \Sigma \quad \longleftrightarrow \quad \left.\frac{\partial_r S}{\partial z^a}\right|_{z^a_0} =0 \,.
\end{equation}
Condition (\ref{properness}) tells us that the rank of the matrix $\frac{\partial_l \partial_r S}{\partial z^a \partial z^b}$ is half its dimensions. Due to (\ref{masterequation}) in the next paragraph, this is the maximum that can be achieved and it guarantees that all the symmetries in the theory have been taken care of via the introduction of ghosts, zero modes and their antifields.
\end{itemize}
If we take into account the boundary conditions, the expansion in (\ref{expansion}) looks like
\begin{eqnarray} \label{actionBV}
S_0&=&S_0[\phi]\,, \nonumber \\
S_1&=&\phi^*_i\RS{}^i{}_{a_0}c_{(0)}{}^{a_0}\,, \nonumber \\\nonumber
S_2&=&c^*_{(0) a_0}\left(\ZSone{}^{a_0}{}_{a_1}c_{(1)}{}^{a_1}+\frac{1}{2}\TS{}^{a_0}{}_{b_0c_0}c_{(0)}{}^{c_0}c_{(0)}{}^{b_0} \right)\\&&+\phi^*_i\phi^*_j\left(\frac{1}{2}\VSone{}^{ji}{}_{a_1}c_{(1)}{}^{a_1}+\frac{1}{4}\ES{}^{ij}{}_{a_0 b_0}c_{(0)}{}^{a_0}c_{(0)}{}^{b_0}\right)\,, \nonumber \\
S_3&=&c^*_{(1) a_1}\left(\ZStwo{}^{a_1}{}_{a_2}c_{(2)}{}^{a_2} +\ldots +
 \frac{1}{2}{\widetilde {\boldsymbol
 F}}{}^{a_1}{}_{e_0d_0b_0}c_{(0)}{}^{b_0}c_{(0)}{}^{d_0}c_{(0)}{}^{e_0}\right)
 \nonumber\\
 && +c^*_{(0)
 a_0}\phi^*_i\left(\VStwo{}^{i a_0}{}_{a_2}c_{(2)}{}^{a_2}
  + \ldots -\frac{1}{2}{\widetilde {\boldsymbol
 D}}{}^{i a_0}{}_{e_0d_0b_0}c_{(0)}{}^{b_0}c_{(0)}{}^{d_0}c_{(0)}{}^{e_0}\right)+\ldots
\end{eqnarray}
Here we have written down the terms up to antifield number 3. The objects $\RS$, $\ZSone$, $\TS$, $\VSone$, $\ES$, $\ZStwo$, ${\widetilde {\boldsymbol
 F}}$, $\VStwo$ and $\widetilde {\boldsymbol
 D}$ should be thought of as generic functionals of the fields $\phi^i$ (not of the ghosts and antifields!) with a particular index structure. The dots in $S_3$ denote more tensors with different index structures, which we do not write explicitly here to clarify the discussion. Also note that each term in (\ref{actionBV}) has ghost number zero and even parity, and that the classical limit (\ref{classicallimit}) is satisfied.

In the next paragraph, we will impose an extra equation on the extended action $S$, which will lead to a particular form for the different tensors.

\paragraph{The classical master equation and general solutions.}
The equation that we will impose is called the classical `master equation' and it takes the form
\begin{equation}
(S,S)=0\,, \label{masterequation}
\end{equation}
where $S$ is the extended action that we introduced in (\ref{expansion}) and that satisfies the boundary conditions (\ref{classicallimit}) and (\ref{properness}). Using (\ref{bosonfunctional}), the master equation can also be written as
\begin{equation}
2\frac{\partial_r S}{\partial \chi^n}\frac{\partial_l S}{\partial \chi^*_n}=0\,.
\end{equation}

To see what this equation really means, we plug in the expansion (\ref{expansion}) into the left hand side of (\ref{masterequation}). We get
\begin{eqnarray} \label{MEexpanded}
(S,S)&=&2\frac{\partial_r S}{\partial \chi^n}\frac{\partial_l S}{\partial \chi^*_n} \nonumber \\
&=&2\frac{\partial S_0}{\partial \phi^j}\RS{}^j_{a_0}c_{(0)}{}^{a_0}+\phi^*_i\left(2\frac{\partial \RS{}^i_{a_0}}{\partial \phi^j}\RS{}^j_{b_0}-\RS{}^i_{c_0}\TS{}^{c_0}{}_{a_0 b_0}+\frac{\partial S_0}{\partial \phi^j}\ES{}^{ji}{}_{a_0 b_0}\right)c_{(0)}{}^{a_0}c_{(0)}{}^{b_0} \nonumber \\
&&+2\phi^*_i\left(\RS{}^i{}_{a_0}\ZSone{}^{a_0}{}_{a_1}-\frac{\partial S_0}{\partial \phi^j}\VSone{}^{ji}{}_{a_1}\right)c_{(1)}{}^{a_1} + \ldots \nonumber \\
&&+2c^*_{(0)a_0}\left(\ZSone{}^{a_0}{}_{a_1}\ZStwo{}^{a_1}{}_{a_2}-\frac{\partial S_0}{\partial \phi^j}\VStwo{}^{j a_0}{}_{a_2}\right)c_{(2)}{}^{a_2}+\ldots
\end{eqnarray}
for the first few terms. To demand that this expression is zero (which is the content of the master equation), means that all the different terms in (\ref{MEexpanded}) should vanish separately.
We see that the vanishing of the first, second and third term is equivalent with equations (\ref{Noether}), (\ref{offshellalgebra}) and (\ref{zeromodesS}) respectively.
So it is clear that the master equation is satisfied up to antifield number 1 when we identify $\RS$, $\TS$, $\ES$, $\ZSone$ and $\VSone$ in (\ref{actionBV}) with
the ones in section \ref{s:closure}. In other words, the master equation demands that the $\RS$ are exactly the gauge generators, the $\TS$ are the structure constants, the $\ZSone$ are the first stage zero modes etc.

This discussion can be continued to terms with higher antifield number. For example, the last line in (\ref{MEexpanded}) vanishes if also the $\ZStwo$ and $\VStwo$ tensors are identified with the ones in section \ref{s:closure}.
Eventually, due to the uniqueness of the solution $S$ (see \cite{Batalin:1985qj, Fisch:1989rp, Voronov:1982cp, Vandoren:1993bw} for a proof), we conclude that the dots in (\ref{MEexpanded}) lead to \textit{all} the relations that determine the gauge structure. For example, at higher order we will also discover the higher order zero modes (\ref{arbitrarystagezeromod}).

To summarize, we have seen that the unique solution $S$ of the master equation (\ref{masterequation}), supplemented by the boundary conditions (\ref{classicallimit}) and (\ref{properness}), is an expansion in the antifields that contains all the gauge structure tensors of the theory as its expansion coefficients. It is in this sense that all the details of the gauge structure of the theory are contained in one equation and that the BV-formalism provides a concise framework for the complicated properties of the gauge algebra.

Before we apply this strong result to our example of the embedding tensor formalism, let us make a final remark about gauge theories without an action.
In section \ref{s:gaugealgebra} we have encountered an example of a consistent gauge algebra that closes on the fields, but without the existence of a Lagrangian description for these fields. Since our discussion on the BV formalism explicitly assumes the existence of a classical action $S_0$, one might wonder whether the case without an action can also be incorporated. This turns out to be possible if one makes the following modifications to the original formulation of the BV formalism. First, we set $S_0=0$, so there is no zeroth order term in the extended action $S$. Due to the absence of $S_0$, the proof of the uniqueness of the solution for S breaks down.\footnote{The Koszul-Tate differential is no more
acyclic.} As a consequence the terms with $\phi^*_i \phi^*_j$ in $S_2$ are undetermined (as well as several other terms at higher order in the antifields).
We can delete these terms, and we find a solution without any terms
quadratic in antifields. In turn this leads to the vanishing of all terms in (\ref{MEexpanded}) that are proportional to the field equations. In other words, $(S,S)$ reduces to
\begin{eqnarray}
(S,S)&=&\phi^*_i\left(2\frac{\partial \R{}^i_{a_0}}{\partial \phi^j}\R{}^j_{b_0}-\R{}^i_{c_0}\T^{c_0}{}_{a_0 b_0}\right)c_{(0)}{}^{a_0}c_{(0)}{}^{b_0} \nonumber \\
&&+2\phi^*_i\R{}^i{}_{a_0}\Zone{}^{a_0}{}_{a_1}c_{(1)}{}^{a_1}+2c^*_{(0)a_0}\Zone{}^{a_0}{}_{a_1}\Ztwo{}^{a_1}{}_{a_2}c_{(2)}{}^{a_2}+\ldots\,,
\end{eqnarray}
where we have also removed all the tildes in order to distinguish these tensors from the ones in the presence of an action. Again, if we impose the master equation, we encounter the relations (\ref{closure}), (\ref{zeromodes}), (\ref{zeromodes2}) etc. which determine all the properties of the gauge structure.

\subsection{Embedding tensor formalism and the BV formulation \label{s:embeddingBV}}
Let us now apply the results form the previous section to the embedding tensor formalism. The field content can easily be identified and is summarized in Table \ref{table2}.
\begin{table}[t]
 \begin{center}\renewcommand{\arraystretch}{1.5}
 \begin{tabular}{|l|lll|c|c|c|}
\cline{5-7}\multicolumn{4}{c|}{}&parity&ghost $\sharp$&antifield $\sharp$\\
    \hline fields $\phi^i$&&$A_\mu{}^M$&$B_{\mu\nu}{}^{\lc MN \rf}$&+&0&0\\
ghosts $c_{(0)}{}^{a_0}$ & $c_{(0)}{}^{M}$&$c_{(0)}{}_\mu{}^{\lc MN\rf}$&$c_{(0)}{}_{\mu\nu}{}^{\lc M\lc NP\rf\rf}$&$-$&1&0\\
$c_{(1)}{}^{a_1}$&$c_{(1)}{}^{\lc MN\rf}$&$c_{(1)}{}_{\mu}{}^{\lc M\lc NP\rf\rf}$&$c_{(1)}{}_{\mu\nu}{}^{\lc M\lc N \lc PQ\rf\rf\rf}$&+&2&0\\
\multicolumn{1}{|c|}{$\vdots$}&\multicolumn{1}{|c}{$\vdots$}&\multicolumn{1}{c}{$\vdots$}&\multicolumn{1}{c}{$\vdots$}&\multicolumn{1}{|c|}{$\vdots$}&\multicolumn{1}{|c|}{$\vdots$}&\multicolumn{1}{|c|}{$\vdots$}
\\\hline
  \end{tabular}\caption{\label{table2} Field content of the BV formalism.}
\end{center}
\end{table}
Each of the fields in this table gets a corresponding antifield $\phi_i^*$, $c_{(0)a_0}^*$, etc. The ghost number, parity and antifield number of the antifields can be determined via (\ref{ghantifield}), (\ref{parantifield}) and (\ref{afn}) respectively.

Then we construct the extended action and impose the master equation. From our considerations of the previous section, we know that $S$ is given by the expansion in (\ref{expansion}). However, the precise form of the $S_i$ depends on whether we consider the embedding tensor formalism in the absence or the presence of an action, i.e. in terms of the untilded or tilded tensors respectively. In the first case, the leading terms in the extended action are
\begin{eqnarray}
  \label{actionBVembedded}S&=&\phi^*_i \R^i{}_{a_0}c_{(0)}{}^{a_0}+c_{(0)}^*{}_{a_0} \Zone{}^{a_0}{}_{a_1}c_{(1)}{}^{a_1}+c^*_{(1)}{}_{a_1}\Ztwo{}^{a_1}{}_{a_2}c_{(2)}{}^{a_2}+\hdots\\\nonumber
  &=&\begin{array}[t]{lllr}
  A^{*\,\mu}{}_M \big{(}&D_\mu{}^M{}_K c_{(0)}{}^K&-Y^M{}_{KL}c_{(0)}{}_\mu{}^{\lc KL\rf}&\big{)}\\
  +B^{*\mu\nu}{}_{\lc MN \rf}\big{(}&-2 {\cal H}_{\mu\nu}{}^{\lc M}c_{(0)}{}^{N\rf}&+2D_{[\mu}{}^{MN}{}_{KL}c_{(0)\,\nu]}{}^{\lc KL\rf}&\\
  &-Y^{MN}{}_{K\lc LR\rf}c_{(0)\,\mu\nu}{}^{\lc K\lc LR\rf\rf}&&\big{)}\\
  +\hdots\\
  +c_{(0)}^*{}_M\big{(}&Y^M{}_{KL}c_{(1)}{}^{\lc KL \rf}&&\big{)}\\
  +c_{(0)}^*{}^\mu{}_{\lc MN\rf}\big{(}&D_\mu{}^{MN}{}_{KL}c_{(1)}{}^{\lc KL \rf}&+Y^{MN}{}_{K\lc LR\rf}c_{(1)}{}_{\mu}{}^{\lc K\lc LR \rf\rf}&\big{)}\\
  +c_{(0)}^*{}^{\mu\nu}{}_{\lc M \lc NP\rf\rf}\big{(}&-{\cal H}_{\mu\nu}{}^{\lc M}{\mathbb P}^{NP\rf}{}_{KL}c_{(1)}{}^{\lc KL \rf}&+2D_{[\mu}{}^{ M\lc NP\rf}{}_{K \lc LR\rf}c_{(1)}{}_{\nu]}{}^{\lc K \lc LR\rf\rf}&\\
  &+Y^{M\lc NP\rf}{}_{K \lc L \lc RS\rf\rf}c_{(1)}{}_{\mu\nu}{}^{\lc K \lc L \lc RS\rf\rf\rf}&&\big{)}\\
  +\hdots\\
  +c_{(1)}^*{}_{\lc MN\rf}\big{(}&-Y^{MN}{}_{K\lc LR\rf}c_{(2)}{}^{\lc K\lc LR\rf\rf}&&\big{)}\\
  +c_{(1)}^*{}^{\mu}{}_{\lc M\lc NP\rf\rf}\big{(}&D_\mu{}^{\lc M\lc NP\rf\rf}{}_{K\lc LR\rf}c_{(2)}{}^{\lc K\lc LR\rf\rf}&-Y^{ M\lc NP\rf}{}_{K\lc L\lc RS\rf\rf}c_{(2)}{}_{\mu\nu}{}^{\lc K\lc L\lc RS\rf\rf\rf}&\big{)}\\
  +\hdots
  \end{array}
\end{eqnarray}
We only wrote down the covariant terms and we recognize a certain systematics in this expression. At each level in the antifields, we encounter the same objects $Y$, $D_\mu$ and ${\cal H}$ between the brackets, but multiplied by different ghost fields. This is due to the particular form of the gauge transformations and (higher order) zero modes. Also for higher levels in the antifields, we expect that this structure survives. The dots in (\ref{actionBVembedded}) denote extra terms that contain non-covariant objects and higher orders in the antifields.

The same calculation can be done for the embedding tensor formalism in the presence of an action. Then the extended action $S$ contains extra terms, starting with the classical action $S_0$ (see \ref{action4d}) at zeroth order. Also other new terms are present at higher orders in the antifields, as can be seen from (\ref{actionBV}). Since we identified the expansion coefficients ($\RS$, $\TS$, $\ES$, etc.) in (\ref{actionBV}) with the gauge structure tensors in section \ref{s:gaugealgebra}, the latter can be substituted into the expressions for $S_i$ in (\ref{actionBV}). We will not do this again, since the final result looks very similar to (\ref{actionBVembedded}).

All in all, we have shown that the BV formalism provides a very appropriate description for the complicated gauge structure of the embedding tensor formalism. It suffices to consider the extended action $S$ and assume the master equation $(S,S)=0$, in order to have a full handle on the gauge structure of the theory. To finish this section, we will further illustrate this by means of an example. We will consider the terms in the master equation that are proportional to $c^*_{(0)\,a_0}c_{(0)}{}^{b_0}c_{(0)}{}^{d_0}c_{(0)}{}^{e_0}$, and show that they give rise to the modified Jacobi identity. We start from
\begin{eqnarray}
\label{exampleBV} (S,S)&=&\hdots + 2\frac{\partial_r S_2}{\partial \phi^i}\frac{\partial_l S_1}{\partial \phi_i^*}+2\frac{\partial_r S_2}{\partial c_{(0)}{}^{a_0}}\frac{\partial_l S_2}{\partial c^*_{(0)}{}_{a_0} }+2\frac{\partial_r S_2}{\partial c_{(1)}{}^{a_1}}\frac{\partial_l S_3}{\partial c^*_{(1)}{}_{a_1} }+\hdots\\\nonumber
&=&c^*_{(0)}{}_{a_0}\left(\frac{\partial \TS{}^{a_0}{}_{b_0 d_0}}{\partial \phi^i}\RS{}^i{}_{e_0}+\TS{}^{a_0}{}_{b_0c_0}\TS{}^{c_0}{}_{d_0e_0}+\ZSone{}^{a_0}{}_{a_1}{\widetilde {\boldsymbol F}}{}^{a_1}{}_{e_0d_0b_0}\right)c_{(0)}{}^{b_0}c_{(0)}{}^{d_0}c_{(0)}{}^{e_0}\\\nonumber
&&+\hdots
\end{eqnarray}
Note that in the absence of a classical action $S_0$ in the embedding tensor formalism, the expression (\ref{exampleBV}) is completely analogous, except that tilded tensors should be replaced by untilded ones.

If the master equation is satisfied, the terms that are proportional to $c^*_{(0)\,a_0}c_{(0)}{}^{b_0}c_{(0)}{}^{d_0}c_{(0)}{}^{e_0}$ should vanish, i.e.
\begin{equation}
c^*_{(0)}{}_{a_0}\left( \frac{\partial \TS{}^{a_0}{}_{b_0 d_0}}{\partial \phi^i}\RS{}^i{}_{e_0}+\TS{}^{a_0}{}_{b_0c_0}\TS{}^{c_0}{}_{d_0e_0}+\ZSone{}^{a_0}{}_{a_1}{\widetilde {\boldsymbol F}}{}^{a_1}{}_{e_0d_0b_0}\right)c_{(0)}{}^{b_0}c_{(0)}{}^{d_0}c_{(0)}{}^{e_0}=0\,.
\end{equation}
This imposes several relations between the gauge generators $\RS$, structure functions $\TS$, zero modes $\ZSone$ and tensors ${\widetilde {\boldsymbol F}}$. Let us calculate the easiest contribution, i.e. for the indices $a_0,\, b_0,\, d_0,\, e_0\, \in\, \{K,\,L,\,M,\,\hdots\}$, and plug in the expressions for the structure functions and zero modes,
\begin{equation}
\label{exampleBV2} c^*_{(0)}{}_{K}\left( \frac{\partial X_{[L M]}{}^{K}}{\partial \phi^i}\RS{}^i{}_{N}+X_{[L P]}{}^{K}X_{[MN]}{}^{P}+Y^{K}{}_{PQ}{\widetilde {\boldsymbol F}}{}^{\lc PQ\rf}{}_{NML}\right)c_{(0)}{}^{L}c_{(0)}{}^{M}c_{(0)}{}^{N}=0\,.
\end{equation}
The first term between the brackets vanishes because the $X_{LM}{}^{K}$ do not depend on the $\phi^i$. The second term is antisymmetric in $[MNL]$ since it is multiplied by the anticommuting ghost fields, and therefore it is equal to the left hand side of the modified Jacobi identity (\ref{Jacobiidentity}). Finally, the third term in (\ref{exampleBV2}) can accommodate the right hand side of the modified Jacobi identity, since it is proportional to $Y^K{}_{PQ}$. If we set
\begin{equation}
 {\widetilde {\boldsymbol F}}{}^{\lc PQ\rf}{}_{NML}c_{(0)}{}^{L}c_{(0)}{}^{M}c_{(0)}{}^{N}=\frac{1}{3} \delta_{N}{}^{\lc P} X_{ML}{}^{Q \rf}c_{(0)}{}^{L}c_{(0)}{}^{M}c_{(0)}{}^{N}\,,
\end{equation}
we have shown that at antifield number $2$ in the master equation, the modified Jacobi identity appears. This is clearly a consequence of the presence of the non-vanishing zero modes, that allow for an extra term that is proportional to $Y^K{}_{PQ}$.
Likewise, several other relations can be found that are a consequence of the existence of the zero modes. Another example is the relation $Y^P{}_{RS} X_{PM}{}^N=0$ that appears if one collects the terms proportional to $\phi_i^* c_{(0)}{}^{a_0}c_{(0)}{}^{b_0}$ in the master equation.

In the end, starting from the extended action and imposing the master equation, we are able to reproduce all the important relations that characterize the gauge structure of the embedding tensor formalism, and that were found before in the literature (e.g. in \cite{deWit:2005ub,deWit:2008ta,deWit:2008gc}).

\section{Conclusions \label{s:conclusions}}
This article extends previous work that has been done on the $D=4$ embedding tensor formalism. It emphasizes the complicated form of the gauge algebra that was previously discussed in \cite{deWit:2008ta, deWit:2008gc}, and tries to suggest a more concise description of the formalism via BV theory.

We started by calculating the full gauge algebra on the 1- and 2-form gauge fields. As these fields suffice to write down a gauge invariant action in 4 dimensions \cite{deWit:2005ub}, no higher order form fields were considered. We argued that the algebra in the absence of any dynamics for the fields explicitly differs from the algebra in the presence of a gauge invariant action.
In the latter case we showed that the algebra is open, i.e. only closes on-shell, whereas in the first case the algebra turned out to be closed. In both cases the algebra is soft since the `structure constants' are functions of the fields. We also calculated the zero modes of the gauge transformations and proved that in both cases the algebra is higher-stage reducible. In principle we could conclude that the embedding tensor formalism is even infinite stage reducible because the level at which the zero modes become independent cannot be determined. But as the discussion was very generic, we suggest that a case-by-case study of particular examples can bring more insight into this.

After having determined the relevant gauge structure tensors (generators, structure constants, zero modes, etc.) we used these tensors to construct a BV action. In this way all the features of the complicated gauge structure are captured by the BV framework and we conclude that this framework can be a convenient tool to further investigate the embedding tensor formalism.

An alternative approach would be to use the BV method of constructing stepwise an extended action, starting from a classical action. If we impose the $(S,S)=0$ condition on the extended action, each term in the expansion must vanish separately and this gives rise to the known gauge structure relations (commutation relations, zero mode relations, Jacobi identities, etc.). Then the properties of the gauge structure tensors that we mentioned in earlier chapters follow from these relations.

It would be interesting to extend our results to generic dimensions (especially in the cases where an action is known) and to study the gauge structure of the full tensor hierarchy, i.e. including higher order $p$-form fields. Also the reducibility of the theory remains an open question. As we said above, our discussion so far was very generic. Studying specific examples for which an explicit form of the projectors $\mathbb P$ is chosen can help us to get a better understanding of the level $L$ at which all zero modes become independent.
Another way to study the reducibility of the theory is to do a dimensional analysis of the degrees of freedom of the theory. The total number of degrees of freedom of the theory depends strongly on the number of (higher stage) zero modes. By calculating this number explicitly, we can determine the level of reducibility $L$. This calculation might even be possible for generic models and arbitrary spacetime dimensions $D$.

Another subject to look at in the future is the quantization of generic gauge theories. In this article we exploited the fact that the BV formalism provides a compact notation for the gauge structure of \textit{classical} theories. On the other hand, the BV formalism was originally designed as a method for the \textit{quantization} of field theories. So, due to our reformulation of the embedding tensor formalism in terms of the BV formalism, we have now all the tools available for the quantization of generic gauged supergravities. In practice however, this might still be very hard to do.

\section*{Acknowledgements}
We would like to thank H. Samtleben and J. Hartong for valuable discussions.
This work is supported in part by the FWO - Vlaanderen, project G.0235.05
and in part by the Federal Office for Scientific, Technical and Cultural Affairs
through the ‘Interuniversity Attraction Poles Programme -– Belgian Science
Policy’ P6/11-P. J.D.R. is an Aspirant of the FWO-Vlaanderen.

\newpage
\appendix
\section{Useful relations \label{a:usefulrel}}

\subsection{Covariant and contravariant tensors \label{a:covcontra}}
We call an object $T^M$ \textit{contravariant} if it transforms as follows under the gauge group
\begin{equation}
\delta(\Lambda) \, T^M =\Lambda^K\delta_K T^M=-\Lambda^KX_{KN}{}^MT^N.
\end{equation}
As an example, $T^M$ can be thought of as the modified field strength ${\cal H}_{\mu\nu}{}^M$, for which the transformation is given in (\ref{deltaH}).

We call an object $T_M$ \textit{covariant} if its gauge transformations are
\begin{equation}
\delta(\Lambda) \, T_M =\Lambda^K\delta_K T_M=\Lambda^KX_{KM}{}^NT_N.
\end{equation}
These transformations can be trivially generalized to objects with an arbitrary number of upper and lower vector indices.

However, in section \ref{s:structure} we encountered the special case where tensors $T^{RS}$ are multiplied by $Y^M{}_{RS}$. Generically, the latter does not map onto the full symmetric tensor product $(RS)$. To make this more precise, we introduced
a projector (\ref{defP2}) which defines a restricted representation denoted by the brackets
$\lc \hdots \rf$ in (\ref{specialbrackets}). Then $T^{\lc RS \rf}$ and $ T_{\lc RS \rf}$ are objects that transform as follows under the restricted representation
\begin{eqnarray} \label{deltaT2}
 \delta(\Lambda)\, T^{\lc RS \rf} &\equiv& -\Lambda^K X_{K \lc LM \rf}{}^{\lc RS \rf} T^{\lc LM \rf}\,,\\
\delta(\Lambda)\, T_{\lc RS \rf} &\equiv&  \Lambda^K X_{K \lc RS \rf}{}^{\lc LM \rf} T_{\lc LM \rf}\,.
\end{eqnarray}
This defines a new tensor $X_{K \lc LM \rf}{}^{\lc RS \rf}$.
Since the gauge transformation of $T^{\lc RS \rf}$ can also be written as
\begin{equation}
 \delta_K T^{\lc RS \rf}={\mathbb P}^{RS}{}_{MN}\left(-X_{KL}{}^{M}T^{\lc LN\rf}-X_{KL}{}^{N}T^{\lc ML \rf}\right)\,,
\end{equation}
we conclude that
\begin{equation}
 X_{K \lc LM \rf}{}^{\lc RS \rf} = 2X_{K \lc L}{}^{\lc R}\delta_{M \rf}^{S \rf}\,.
\end{equation}

With this tensor we can build a new tensor $Y^{MN}{}_{P \lc RS \rf}$:
\begin{equation} \label{Y2}
Y^{MN}{}_{P \lc RS \rf} \equiv 2\delta_P{}^{\lc M}Y^{N \rf}{}_{RS}-X_{P \lc RS \rf}{}^{\lc MN \rf},
\end{equation}
for which $Y^K{}_{MN}Y^{MN}{}_{P \lc RS \rf}=0$ (we will prove this in \ref{a:userel}). Note that the definition (\ref{Y2}) is consistent with the one in (\ref{defY2}).
Generically, the tensor $Y^{MN}{}_{P \lc RS \rf}$ does not map onto the full tensor product $P \lc RS \rf$ but
only on a restricted subrepresentation. Again, we can define a projector\footnote{In this text we only use the properties $\mathbb{P}^{K \lc LM \rf}{}_{P \lc [RS] \rf}=0$ and $\mathbb{P}^{K \lc LM \rf}{}_{P \lc RS \rf}\mid_{(PRS)}=0$. Again $\mathbb{P}^{K \lc LM \rf}{}_{P \lc RS \rf}$ can be taken of lower rank in particular applications where the constraints are satisfied, similar to footnote \ref{fn:projector}.} $\mathbb{P}^{K \lc LM \rf}{}_{P \lc RS \rf}$ such that it leaves the $Y$-tensor invariant:
\begin{equation}
Y^{NQ}{}_{P \lc RS \rf}=Y^{NQ}{}_{K \lc LM \rf}\mathbb{P}^{K \lc LM \rf}{}_{P \lc RS \rf}\,.
\end{equation}
With this new projector, we construct objects that transform in the restricted representation:
\begin{equation}
T^{\lc P\lc RS \rf \rf}\equiv\mathbb{P}^{P \lc RS \rf}{}_{K \lc LM \rf}T^{ K\lc LM \rf }\,.
\end{equation}
And analogous to (\ref{deltaT2}), we introduce a tensor $X_{P \lc Q \lc RS \rf \rf}{}^{\lc M \lc NK \rf \rf}$ that is defined via this gauge transformation:
\begin{equation} \label{deltaT3}
 \delta_K T^{\lc P\lc RS \rf \rf} \equiv -X_{K \lc L \lc MN \rf \rf}{}^{\lc P \lc RS \rf \rf} T^{\lc L \lc MN \rf \rf}.
\end{equation}
This $X_{K \lc L \lc MN \rf \rf}{}^{\lc P \lc RS \rf \rf}$ appears again in a new $Y$-tensor
\begin{equation}\label{defY3}
 Y^{P \lc RS \rf}{}_{K \lc L \lc MN \rf \rf} \equiv -\delta_{K}{}^{\lc P}Y^{RS \rf}{}_{L \lc MN \rf}-X_{K \lc L \lc MN \rf \rf}{}^{\lc P \lc RS \rf \rf}
\end{equation}
for which $Y^{VW}{}_{P \lc RS \rf}Y^{P \lc RS \rf}{}_{K \lc L \lc MN \rf \rf}=0$ (we will prove this in the \ref{a:userel}).

Then we see that there exists a whole hierarchy of such tensors that are mutually orthogonal. A generic $Y$-tensor is defined as
\begin{equation} \label{Yp}
 Y^{M_1 \lc M_2 \lc \ldots M_p \rf .. \rf}{}_{N_0 \lc N_1 \lc \ldots N_p \rf .. \rf}=-\delta_{N_0}{}^{\lc M_1}Y^{M_2 \lc \ldots M_p \rf .. \rf}{}_{N_1 \lc N_2 \lc \ldots N_p \rf .. \rf}-X_{N_0 \lc N_1 \lc \ldots N_p \rf .. \rf}{}^{\lc M_1 \lc M_2 \lc \ldots M_p \rf .. \rf},
\end{equation}
for $p \geq 3$ and $X_{N_0 \lc N_1 \lc \ldots N_p \rf .. \rf}{}^{\lc M_1 \lc M_2 \lc \ldots M_p \rf .. \rf}$ determines the gauge transformation of a covariant object $T^{\lc M_1 \lc M_2 \lc \ldots M_p \rf .. \rf}$:
\begin{equation} \label{deltaTp}
 \delta_{N_0} T^{\lc M_1 \lc M_2 \lc \ldots M_p \rf .. \rf} \equiv -X_{N_0 \lc N_1 \lc \ldots N_p \rf .. \rf}{}^{\lc M_1 \lc M_2 \lc \ldots M_p \rf .. \rf} T^{\lc N_1 \lc \ldots N_p \rf .. \rf}.
\end{equation}
In the remaining part of this appendix, we will prove some useful relations for the $Y$-tensors that are used throughout the text.

\subsection{Orthogonality of the $Y$-tensors\label{a:userel}}
We will prove that
\begin{equation} \label{YY1}
Y^V{}_{RS}Y^{RS}{}_{K \lc LM \rf}=0
\end{equation}
and
\begin{equation} \label{YY2}
Y^{VW}{}_{P \lc RS \rf}Y^{P \lc RS \rf}{}_{K \lc L \lc MN \rf \rf}=0\,.
\end{equation}
Knowing this, it is then easy to show (by induction) that
\begin{equation} \label{YYp}
 Y^{K_2 \lc K_3 \lc \ldots K_p \rf .. \rf}{}_{M_1 \lc M_2 \lc \ldots M_p \rf .. \rf}Y^{M_1 \lc M_2 \lc \ldots M_p \rf .. \rf}{}_{N_0 \lc N_1 \lc \ldots N_p \rf .. \rf}=0,
\end{equation}
for $p > 3$.

The closure constraint (\ref{closureconstr}) tells us that the embedding tensor is gauge invariant and thus $\delta_V Y^M{}_{RS}=0$. From (\ref{deltaT2}) we know that
\begin{eqnarray}
0&=&\delta_K Y^V{}_{LM} \nonumber \\
 &=&-X_{KR}{}^V Y^R{}_{LM}+X_{K\lc LM \rf}{}^{\lc RS \rf}Y^V{}_{RS} \nonumber \\
 &=&-2Y^V{}_{KR}Y^R{}_{LM}+X_{K\lc LM \rf}{}^{\lc RS \rf}Y^V{}_{RS}\,, \label{deltaY1}
\end{eqnarray}
where we used (\ref{ZTheta}) in the third equality. Now we can write
\begin{eqnarray}
 Y^V{}_{RS}Y^{RS}{}_{K \lc LM \rf}&=&2Y^V{}_{KR}Y^R{}_{LM}-X_{K\lc LM \rf}{}^{\lc RS \rf}Y^V{}_{RS} \nonumber \\
				  &=&0\,,
\end{eqnarray}
due to (\ref{Y2}) in the first equality and (\ref{deltaY1}) in the second. This proves equation (\ref{YY1}).
A similar computation can be done in order to show (\ref{YY2}). From the closure constraint we conclude that (using (\ref{deltaT3}))
\begin{eqnarray}
0&=&\delta_K Y^{VW}{}_{L \lc MN \rf} \nonumber \\
 &=&X_{K \lc RS \rf}{}^{\lc VW \rf} Y^{RS}{}_{L \lc MN \rf}-X_{K\lc L \lc MN \rf \rf}{}^{\lc R \lc ST \rf \rf}Y^{VW}{}_{R \lc ST \rf}. \label{deltaY2}
\end{eqnarray}
If we combine this expression with (\ref{Y2}) and (\ref{YY1}) it is easy to see that (\ref{YY2}) is satisfied.

Now we will prove the orthogonality (\ref{YYp}) for general $Y$-tensors by induction.
We start from the observation that the $Y$-tensor with $p-1$ upper indices and $p$ lower indices is gauge invariant:
\begin{eqnarray} \label{deltaYp}
X_{N_0 \lc K_2 \lc K_3 \lc \ldots K_p \rf .. \rf}{}^{\lc M_2 \lc M_3 \lc \ldots M_p \rf .. \rf}Y^{ K_2 \lc K_3 \lc \ldots K_p \rf .. \rf}{}_{N_1 \lc N_2 \lc \ldots N_p \rf .. \rf}&=& \nonumber \\
X_{N_0 \lc N_1 \lc \ldots N_p \rf .. \rf}{}^{\lc K_1 \lc K_2 \lc \ldots K_p \rf .. \rf}Y^{M_2 \lc M_3 \lc \ldots M_p \rf .. \rf}{}_{K_1 \lc K_2 \lc \ldots K_p \rf .. \rf}.&&
\end{eqnarray}
This relation allows us to rewrite the lefthandside of equation (\ref{YYp}) as
\begin{eqnarray}
\mbox{(\ref{YYp})}&=&-Y^{K_2 \lc K_3 \lc \ldots K_p \rf .. \rf}{}_{N_0 \lc M_2 \lc \ldots M_p \rf .. \rf}Y^{M_2 \lc M_3 \lc \ldots M_p \rf .. \rf}{}_{N_1 \lc N_2 \lc \ldots N_p \rf .. \rf} \nonumber \\
        \label{} & &-Y^{M_2 \lc M_3 \lc \ldots M_p \rf .. \rf}{}_{N_1 \lc N_2 \lc \ldots N_p \rf .. \rf}X_{N_0 \lc M_2 \lc M_3 \lc \ldots M_p \rf .. \rf}{}^{\lc K_2 \lc K_3 \lc \ldots K_p \rf .. \rf} \,.
\end{eqnarray}
If we now replace the $X$-tensor on the second line by $Y$ tensors via their definition (\ref{Yp}), then this relation reduces to (\ref{YYp}) for $p-2$ upper and $p-1$ lower indices. By induction, this orthogonality relation is satisfied and we have proven (\ref{YYp}).

\subsection{Covariant derivatives \label{a:defrel}}
It is useful to introduce derivative operators $D_{\mu}{}^{N_1 \lc N_2 \lc \ldots N_p \rf .. \rf}{}_{M_1 \lc M_2 \lc \ldots M_p \rf .. \rf}$ that describe
the action of a covariant derivative on objects $T^{\lc M_1 \lc M_2 \lc \ldots M_p \rf .. \rf}$ in different representations of the gauge group. For $p=1$ we define
\begin{equation}
D_{\mu}{}^N{}_M T^M \equiv \left(\delta_M{}^N\partial_{\mu}+A_{\mu}{}^QX_{QM}{}^N\right)T^M=D_{\mu}T^N.
\end{equation}
We extend this definition for generic $p$ as follows
\begin{equation} \label{specder}
 D_{\mu}{}^{N_1 \lc N_2 \lc \ldots N_p \rf .. \rf}{}_{M_1 \lc M_2 \lc \ldots M_p \rf .. \rf} \equiv \partial_{\mu}\mathbb{P}^{N_1 \lc N_2 \lc \ldots N_p \rf .. \rf}{}_{M_1 \lc M_2 \lc \ldots M_p \rf .. \rf} +
																      A_{\mu}{}^Q X_{Q \lc M_1 \lc M_2 \lc \ldots M_p \rf .. \rf}{}^{\lc N_1 \lc N_2 \lc \ldots N_p \rf .. \rf}.
\end{equation}
The following relation is being used in the text
\begin{equation} \label{DY1}
 D_{\mu}{}^N{}_M Y^M{}_{RS}=Y^N{}_{KL}D_{\mu}{}^{KL}{}_{RS}.
\end{equation}
It is not difficult to prove this:
\begin{eqnarray} \label{DYp}
Y^N{}_{KL}D_{\mu}{}^{KL}{}_{RS}&=&Y^N{}_{KL}\left(\partial_{\mu}\mathbb{P}^{KL}{}_{RS}+A_{\mu}{}^QX_{Q \lc RS \rf}{}^{\lc KL \rf}\right) \nonumber \\
		               &=&\partial_{\mu}Y^N{}_{RS}+A_{\mu}{}^QX_{Q M}{}^{N}Y^M{}_{RS} \nonumber \\
			       &=&D_{\mu}{}^N{}_M Y^M{}_{RS},
\end{eqnarray}
where in the second equality we used the definition of the projector $\mathbb{P}^{KL}{}_{RS}$ and the gauge invariance of the $Y^M{}_{RS}$ tensor (\ref{deltaY1}).
Finally, (\ref{DY1}) can be generalized to
\begin{eqnarray}
&D_{\mu}{}^{M_1 \lc M_2 \lc \ldots M_p \rf .. \rf}{}_{K_1 \lc K_2 \lc \ldots K_p \rf .. \rf}Y^{K_1 \lc K_2 \lc \ldots K_p \rf .. \rf}{}_{N_0 \lc N_1 \lc \ldots N_p \rf .. \rf}=& \nonumber \\
&Y^{M_1 \lc M_2 \lc \ldots M_p \rf .. \rf}{}_{K_0 \lc K_1 \lc \ldots K_p \rf .. \rf}D_{\mu}{}^{K_0 \lc K_1 \lc \ldots K_p \rf .. \rf}{}_{N_0 \lc N_1 \lc \ldots N_p \rf .. \rf}\,,&
\end{eqnarray}
which is easy to proof if one uses the gauge invariance of the $Y$-tensors (\ref{deltaYp}).

\newpage
\providecommand{\href}[2]{#2}\begingroup\raggedright
\endgroup

\end{document}